\documentstyle[epsfig,amssymb,floatflt,float,here]{mn2e}

\newif\ifAMStwofonts

\newcommand{\sm}{\, {\rm M}_{\odot}}

\newcommand{\kms}{km~s$^{-1}$}

\newcommand{\mnras}{MNRAS}
\newcommand{\apj}{ApJ}
\newcommand{\aj}{AJ}
\newcommand{\aap}{A\&A}
\newcommand{\nat}{Nature}

\title[Ancient substructure in the Galactic disk]{Pieces of the
puzzle: Ancient substructure in the Galactic disk} \author[A. Helmi et
al.] {Amina Helmi\thanks{E-mail: ahelmi@astro.rug.nl}$^1$, 
J. F. Navarro\thanks{CIAR and Guggenheim Fellow}$^{2,3}$, 
B. Nordstr{\"o}m$^{4,5}$, J. Holmberg$^{6}$, \newauthor
M. G. Abadi$^{2}$\thanks{CITA National Fellow, on leave from 
Observatorio Astron\'omico de C\'ordoba and CONICET, Argentina.} and
M. Steinmetz$^{7,8}$\thanks{David and Lucile Packard Fellow.} \\ \\
$^1$Kapteyn Astronomical Institute, University of Groningen, P.O.Box 800, 
9700 AV Groningen, The Netherlands\\ 
$^2$Department of Physics and Astronomy, University
of Victoria, Victoria, BC V8P 1A1, Canada \\
$^3$Max-Planck Institute for Astrophysics, Karl-Schwarzschild Strasse 1,
D-85741 Garching, Germany\\
$^4$Niels Bohr Institute, Juliane Maries Vej 30, DK-2100 Copenhagen, Denmark,
 and \\ $^5$Lund Observatory, Box 43, SE-22100 Lund, Sweden\\
$^6$Tuorla Observatory, University of Turku, Vais\"al\"antie 20, FIN - 21500 
Piikki\"o, Finland\\
$^{7}$Astrophysikalisches Institut Potsdam, An der Sternwarte 16,
Potsdam 14482, Germany\\
$^{8}$Steward Observatory, University of Arizona, Tucson, AZ 85721, USA\\
}

\pagerange{\pageref{firstpage}--\pageref{lastpage}}
\pubyear{}

\begin{document}

\maketitle

\label{firstpage}

\begin{abstract}

We search for signatures of past accretion events in the Milky Way in
the recently published catalogue by Nordstr\"om et al.~(2004),
containing accurate spatial and kinematic information as well as
metallicities for 13240 nearby stars. To optimize our strategy, we use
numerical simulations and characterize the properties of the debris
from disrupted satellites.  We find that stars with a common
progenitor should show distinct correlations between their orbital
parameters; in particular, between the apocentre A and pericentre P,
as well as their $z$-angular momentum ($L_z$).  In the APL-space, such
stars are expected to cluster around regions of roughly constant
eccentricity.

The APL space for the Nordstr\"om catalogue exhibits a wealth of
substructure, much of which can be linked to dynamical perturbations
induced by spiral arms and the Galactic bar. However, our analysis
also reveals a statistically significant excess of stars on orbits of
common (moderate) eccentricity, analogous to the pattern expected for
merger debris. Besides being dynamically peculiar, the 274 stars in
these substructures have very distinct metallicity and age
distributions, providing further evidence of their extra-Galactic
provenance. It is possible to identify three coherent Groups among
these stars, that, in all likelihood, correspond to the remains of
disrupted satellites.  The most metal-rich group ([Fe/H]~$>-0.45$ dex)
has 120 stars distributed into two stellar populations of $\sim 8$ Gyr
(33\%) and $\sim 12$ Gyr (67\%) of age. The second Group with
$\langle$[Fe/H]$\rangle$~$\sim -0.6$ dex has 86 stars, and shows
evidence of three populations of 8 Gyr (15\%), 12 Gyr (36\%) and 16
Gyr (49\%) of age. Finally, the third Group has 68 stars, with typical
metallicity around $-0.8$ dex, and a single age of $\sim 14$ Gyr. The
identification of substantial amounts of debris in the Galactic disk
whose origin can be traced back to more than one satellite galaxy,
provides evidence of the hierarchical formation of the
Milky Way.
\end{abstract}


\section{Introduction}

In the past two decades, our understanding of galaxy formation has
made a giant leap forward, thanks to both theoretical and
observational developments. We currently know the initial conditions
from which all structure in the Universe formed. The WMAP recent
measurements of the power spectrum of density fluctuations imprinted
on the cosmic microwave background are extremely well fit by a
$\Lambda$CDM cosmogony (Spergel et al. 2003). Therefore, we are in a
good position to study the formation and evolution of galactic systems
within the $\Lambda$CDM concordance model (e.g. Mo, Mao \& White 1998;
Silk \& Bouwens 2001). In this framework, galaxy formation proceeds
hierarchically, from the collapse of small galactic subunits, which
subsequently merge to produce large galaxies like our own (Benson et
al. 2003; Springel \& Hernquist 2003).

It has recently become possible to simulate the formation of a disk
galaxy from first principles in a cosmological context (e.g. Bekki \&
Chiba 2001; Abadi et al. 2003a, Sommer-Larsen, G\"otz \& Portinari
2003; Governato et al. 2004; Brook et al. 2004). Although the
numerical resolution of these simulations (particularly the number of
stellar particles) is far from ideal to study the detailed spatial and
kinematic distribution of stars, they have provided us with useful
insight into the formation process of a disk galaxy (see also Samland
\& Gerhard 2003). In particular, Abadi et al. (2003b) claim that a
majority of the oldest stars in the disk, and the thick disk itself,
originated in satellites accreted on low inclination orbits. However,
it is hard to establish whether this is a generic feature of
hierarchical models because these conclusions were based on a single
cosmological simulation.

In the context of the CDM framework, one would like to determine the
merging history of the Milky Way. How many progenitor systems built up
the Galaxy? What were their properties? What fraction of the stars
located in a region like the Solar neighbourhood formed in situ, what
fraction originated in mergers?  Have major mergers of gas rich
systems played an important role? Is the contribution of minor mergers
significant?  Although the hierarchical model may be fundamentally
different from the collapse model proposed by Eggen, Lynden-Bell \&
Sandage (1962), they may have a few features in common. It is not
unlikely that, early on in the history of the Galaxy, there was a
phase which could be described as a fast dissipative collapse, but
which was triggered by gas-rich mergers (Brook et al. 2004), rather
than by the collapse of a giant cloud of gas. Nevertheless, it is also
clear that collisionless minor mergers have also played a role in the
formation of our Galaxy. This is evidenced by the presence of the
disrupting Sgr dwarf (Ibata, Gilmore \& Irwin 1994), the substructure
in the Galactic halo (Helmi et al. 1999) and the (thick) disk
(i.e. the Arcturus stream, Eggen 1996; Navarro, Helmi \& Freeman
2004), the ring in the outer Galaxy (Yanny et al. 2003), the Canis
major dwarf (Martin et al. 2004; but see also Momany et
al. 2004). This list of examples also serves to indicate that
substructure appears to be ubiquitous, although the samples are still
rather small to establish, even in a statistical sense, how important
mergers have been in the history of the Galaxy.

The spatial and kinematic distribution of the debris from the accreted
satellites is an aspect that has received more attention in the last
few years (Johnston, Hernquist \& Bolte 1996; Helmi \& White 1999;
Helmi, White \& Springel 2003).  For example, Helmi et al. (2003)
using a combination of analytic work and cosmological simulations,
concluded that there should be several hundred kinematically cold
streams in the stellar halo near the Sun. This prediction is
consistent with observations, while a full test would require an
enlargement by a factor ten or so in the size of the samples of halo
stars with accurate 3D kinematics. Nevertheless, Gould (2003) was able
to infer statistically a limit on the granularity of the velocity
distribution of halo stars, using a large sample of nearby dwarfs with
proper motions. He confirmed that there are no streams with more than
5\% of the local density (consistent with the result of Helmi et
al. 1999), and that there are at least 400 such streams present.

However, mergers and accretion are not the only source of substructure
in a galactic disk. Conceivably, these could be the mechanisms that
provide the least prominent contributions to the substructure observed
in disks. Stars form generally in groups or clusters (Chereul,
Cr\'ez\'e \& Bienayme 1998), spiral arms produce perturbations and
resonances that confine stars temporarily to a small region of
phase-space (De Simone, Wu \& Tremaine 2004), and the bar causes
resonances which trap stars into structures such as the Hercules
stream (Dehnen 1988, 2000; Fux 2001). Disentangling the various
physical mechanisms responsible for the substructure observed is by no
means trivial, and requires additional knowledge of the properties of
the members of a given substructure. For example, the age and, in
particular, the metallicity distributions of the Hercules stream are
very broad, --from $-0.6$ dex to $+0.6$ dex peaking around [Fe/H]
$\sim 0$ dex, and with ages in the range 6 to 10 Gyr -- and hence its
origin cannot be attributed to an accreted satellite (Raboud et
al. 1998). It is not inconceivable however, that streams with a merger
origin do show a spread in ages and metallicity, particularly if the
parent system was a relatively large satellite (see e.g. Navarro et
al. 2004).

Interesting constraints on the process of galaxy formation can also be
obtained from detailed analysis of the chemical abundances of stars
(e.g. Cayrel et al. 2004). Recently, Venn et al. (2004) pointed out
that the chemical composition (metallicity and abundance patterns) and
age distribution of stars in the satellites of the Milky Way are
significantly different from those found for stars in the halo and
thin and thick disks of our Galaxy (see also Unavane, Gilmore \& Wyse
1996). They concluded that these components cannot have accreted too
many satellites like those currently orbiting the Milky Way. However,
hierarchical models predict that any accreted population of stars is
more likely to have originated in a few massive galaxies rather than
in many dwarf galaxies. Furthermore, it remains to be seen if the
progenitors of the present-day satellites could have, for example,
built up the Galactic halo, if accreted at high-redshift (as also
suggested by Robertson et al. 2005). This hypothesis can be tested
with accurate age determinations and abundance measurements of the
oldest stellar populations in the dwarf galaxies and in the Galactic
halo (Hill et al. 2005, in preparation).

In this paper, we analyse a recently published catalogue of stars in
the Solar neighbourhood (Nordstr\"om et al. 2004; hereafter N04),
containing accurate full phase-space information and metallicities
([Fe/H]) for 13240 stars.  This sample is therefore ideal to look for
substructures associated with past mergers. We begin our quest in
Section \ref{sec:sample} by describing the sample's properties. To
guide our search for the signatures of disrupted galaxies, in Section
\ref{sec:all_sims} we characterize the properties of substructure
induced by mergers, while we devote Section \ref{sec:analysis} to
study the phase-space distribution of the stars in the N04 sample. In
Sections \ref{sec:apl}, \ref{sec:mc} and \ref{sec:stats} we establish
the presence of substructures in the sample, and study their
properties in Sections \ref{sec:gral} and \ref{sec:satellites}. We
summarise our results in Section \ref{sec:concl}.
\begin{figure*}
\vspace*{0.6cm}
\includegraphics[width=17.5cm,clip]{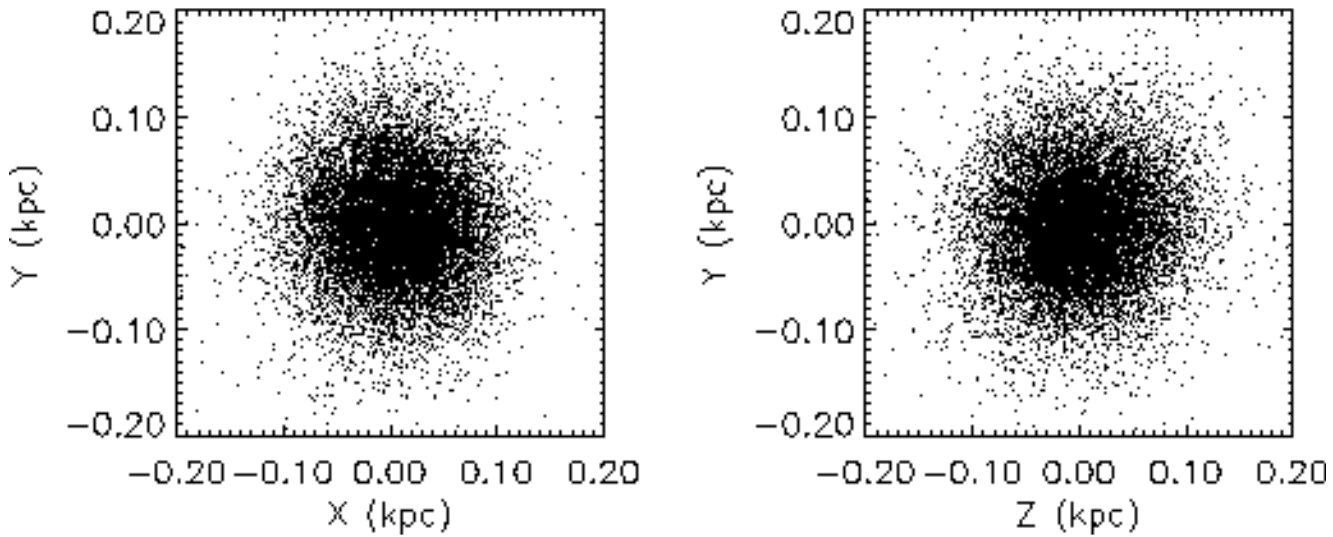}
\caption{Spatial distribution of the stars in the Nordstr\"om et
al. (2004) sample. Note that the sample is restricted to a small
region around the Sun, being volume complete out to a distance of
approximately 40 pc.}
\label{fig:space}
\end{figure*}

\begin{figure*}
\hspace*{-0.5cm}\vspace*{0.4cm}
\includegraphics[width=17.5cm,clip]{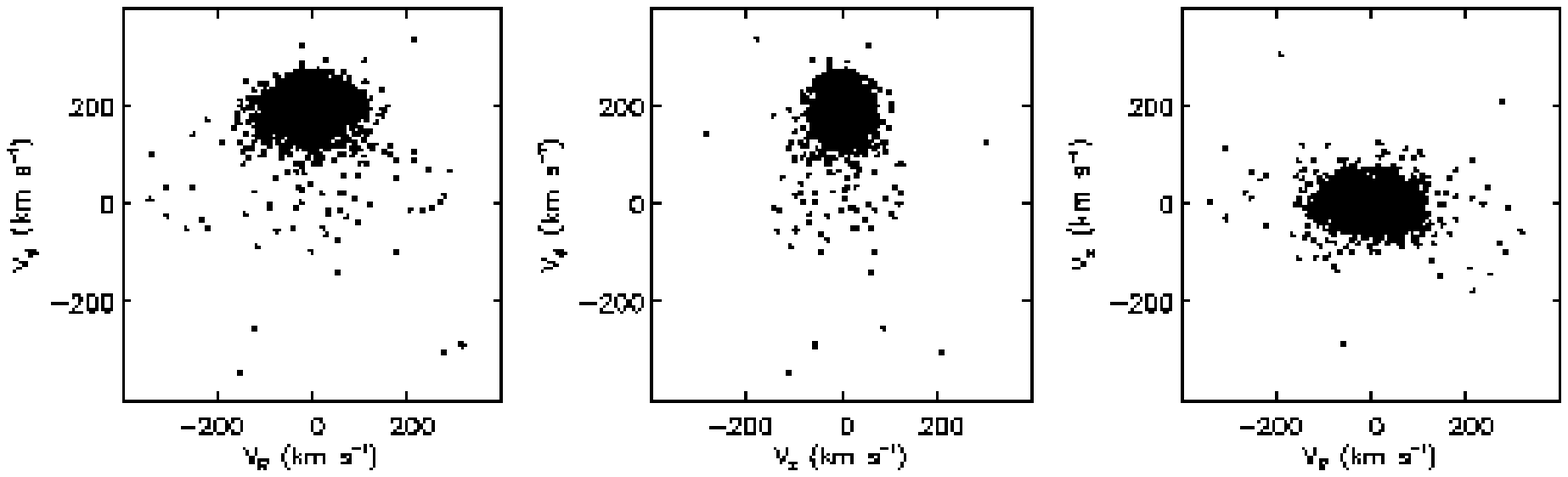}
\vspace*{0.4cm}
\includegraphics[width=17.5cm,clip]{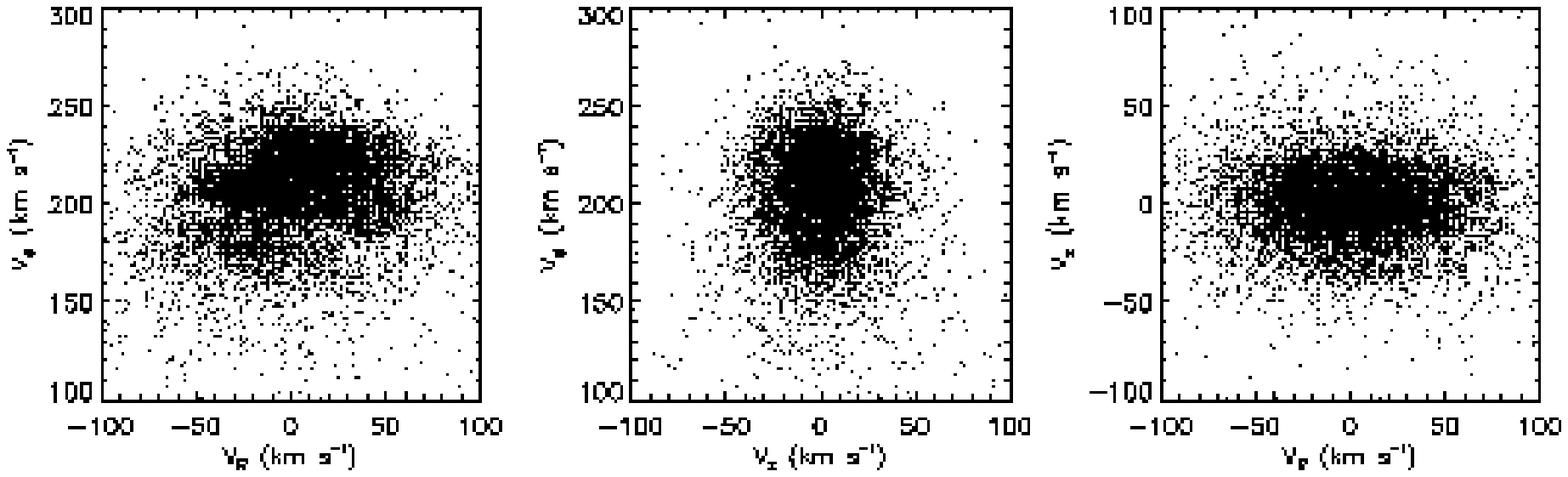}
\vspace*{0.4cm}
\includegraphics[width=17.5cm,clip]{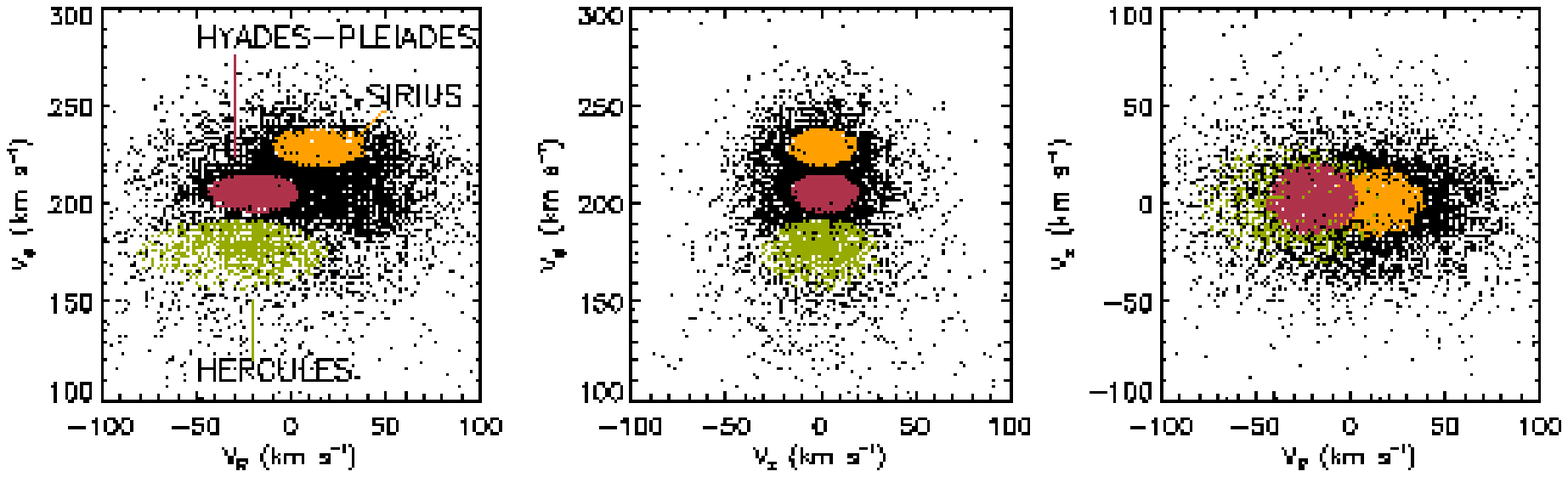}
\caption{{\it Top:} Velocity distribution of stars in the N04
sample. The great majority of the stars have disk-like kinematics. {\it
Centre:} Zoom-in of the top panels to show with higher resolution the
velocity distribution of stars in the Galactic disk. It is immediately
apparent that the kinematic distribution contains a large degree of
substructure. Bottom: Identification of the Hyades-Pleiades, Sirius
and Hercules superclusters in our sample according to the criteria of
Famaey et al. (2005). These are the most prominent structures in
velocity space.}
\label{fig:vel_df}
\end{figure*}

\section{Observational sample}
\label{sec:sample}

In this paper, we focus on the complete, all-sky, magnitude-limited,
and kinematically unbiased sample of 14139\footnote{Full 6D
phase-space information is available for 13240 stars.} F and G dwarf
stars presented by N04. New accurate radial-velocity observations have
been obtained by these authors which, together with published
Str{\"o}mgren $uvby\beta$ photometry, Hipparcos parallaxes, Tycho-2
proper motions, and a few earlier radial velocities, complete the
kinematic information. These high-quality velocity data are
supplemented by effective temperatures and metallicities newly derived
from recent and/or revised calibrations.  Spectroscopic binaries have
been identified thanks to multi-epoch radial velocity measurements.

F- and G-type dwarf stars are particularly useful to study Galactic
evolution, being numerous and long-lived; their atmospheres reflect
their initial chemical composition; and their ages can be estimated
for at least the more evolved stars by comparison with stellar
evolution models. We will now summarise some important properties of
the N04 sample, and refer the reader to the original paper for a more
comprehensive description.

Stars in the N04 sample were selected from a compilation of catalogues
available in the literature with $uvby\beta$ photometry of nearby F
and G stars, but mostly from the surveys by Olsen (1983, 1993, 1994a,
1994b). The N04 sample is volume complete to roughly 40 pc. It is
magnitude complete to $V \le 7.7$ for the bluest stars (slightly
fainter for the reddest G stars), and has a cutoff magnitude $V_{\rm
cut} \sim 8.9$ (and 9.9 for the reddest G stars). The radial
velocities are typically the result of two or more spectroscopic
observations, and have a typical mean error of 0.5 \kms or less. The
proper motions for the vast majority of the stars are from the Tycho-2
catalogue (H{\o}g et al. 2000), i.e.  constructed by combining the
Tycho star-mapper measurements of the Hipparcos satellite with the
Astrographic Catalogue based on measurements in the Carte du Ciel and
other ground-based catalogues. A small number of stars have only one
measurement, either from Hipparcos or Tycho. The typical mean error in
the total proper motion vector is 1.8 mas/yr.

Accurate trigonometric parallaxes are available from Hipparcos for the
majority of the stars (ESA 1997), with relative errors
($\sigma_\pi/\pi$) generally better than 10\%. When the Hipparcos
parallax is either unavailable or is less accurate than the
photometric parallax, whose uncertainty is only 13 \%, the latter is
used. Photometric parallaxes are derived from the distance
calibrations for F and G dwarfs by Crawford (1975) and Olsen (1984).

The accurate determination of metallicities for F and G stars is one
of the strengths of the Str{\"o}mgren $uvby\beta$ system. The
calibrations used in the N04 paper were either already available
(e.g. Schuster \& Nissen 1989), or have been extended towards redder
stars by comparison to spectroscopic metallicities from recently
published high resolution data. The typical uncertainty is of the
order of 0.1 dex. This estimate relies, for example, on the excellent
agreement with the Edvardsson et al. (1993) spectroscopic
metallicities.

Individual ages are determined by means of a Bayesian estimation
method that compares a grid of theoretical isochrones with the
location of stars in the HR diagram, using the observed $T_{eff}$,
$M_V$, and [Fe/H] as well as their observational errors (J{\o}rgensen
and Lindegren 2005). In practice, isochrone ages can only be
determined for stars that have evolved significantly away from the
ZAMS. This precludes the determination of reliable isochrone ages for
unevolved (i.e.  relatively young) stars, and also for low-mass G and
K dwarfs located on the ZAMS.

Figure \ref{fig:space} shows the spatial distribution of the stars in
the sample. Except for the concentration corresponding to the Hyades
(at $x \sim -40$ pc, $y \sim 0$ pc, Perryman et al. 1998), the
distribution of stars in space appears to be smooth.

In Figure \ref{fig:vel_df} we show the velocity distribution of the
stars in the sample. The velocities are defined in a right-handed
Galactic coordinate system with $U$ pointing towards the Galactic
centre, $V$ in the direction of rotation, and $W$ towards the north
Galactic pole. The Sun is located at $x = -8$ kpc, and the velocities
have been corrected for the Solar motion using the values from Dehnen
\& Binney (1998), and assuming a Local Standard of Rest velocity of
220 \kms. It is evident from the top panels of Figure \ref{fig:vel_df}
that most of the stars have a mean prograde motion, indicative of the
dominant contribution of the Galactic thin disk to the current
sample. There are few halo stars present, as evidenced by the small
number of objects that cluster around zero rotational velocity and
that have large radial and $z$-motions. The central panel of Figure
\ref{fig:vel_df} focuses on the stars with disk-like kinematics. Large
amounts of substructure are clearly present in the sample. The most
prominent, and which can be easily picked up by eye, are the
Hyades-Pleiades, Sirius and Hercules superclusters (Eggen 1958, 1960,
1975). Following the characterization by Famaey et al. (2005), we
highlight these structures in the bottom panels of
Fig.~\ref{fig:vel_df}. There are 2103 stars associated with the
Hyades-Pleiades structure, 1353 stars with the Sirius moving group,
and 1071 stars with the Hercules stream in the N04 sample.  These
substructures are most likely associated to two phenomena unrelated to
the merger history of the Galaxy: They are clusters of stars born in
the Galactic thin disk (e.g. Hyades), or they are the result of
perturbations by spiral arms or the Galactic bar which drive the
dynamical evolution of the disk (e.g. Hercules stream).

\begin{figure}
\includegraphics[width=8cm]{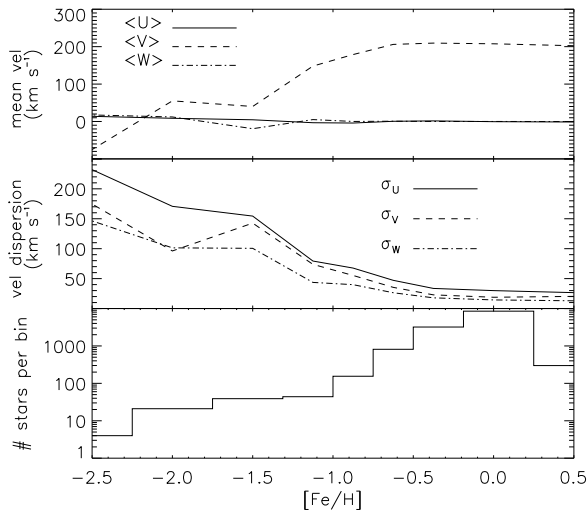}
\caption{Mean velocity components and dispersion as function of
metallicity [Fe/H] for the stars in the N04 sample. The bottom panel
shows that there are only a handful of objects in the most metal-poor
bins, which makes less reliable the characterization of the velocity
ellipsoid.}
\label{fig:mc}
\end{figure}

Almost 99\% of the stars in the sample have accurate metallicities,
which allows us to study how the mean velocity and the dispersion
changes with [Fe/H]. This is illustrated in Figure~\ref{fig:mc} which
shows the mean velocity components and velocity dispersions in the
($U,V,W$) directions as function of [Fe/H].  As extensively discussed
in the literature (Sandage \& Fouts 1987; Norris \& Ryan 1989; Carney
et al. 1990; Nissen \& Schuster 1991; Edvardsson et al. 1993; Chiba \&
Beers 2001), and in particular, in the pioneering work of Eggen et
al. (1962), there is a strong correlation between the kinematics and
the metallicities of stars in the Solar neighbourhood. In broad terms,
the most-metal poor stars form part of a non-rotating, hot component
(halo), while the majority of the most-metal rich stars are in a fast
rotating, cold component (disk).

\section{Insights from numerical simulations}
\label{sec:all_sims}

\subsection{The space of conserved quantities}
\label{sec:intro-apl}

The substructure associated with past accretion events is expected to
have much lower density contrast than, for example, that associated
with open clusters (e.g. Helmi \& White 1999). This implies that
sophisticated methods of identification need to be developed. For
example, satellite galaxies of the size of the Small Magellanic Cloud
are expected to contribute of the order of 50 stellar streams to the
halo in the Solar neighbourhood. Such streams would have rather low
density contrast, and therefore, identification of each one of them
independently could be a rather daunting task. A better strategy is to
identify a space in which all stars from such a satellite are
clustered into a relatively small volume.

Helmi \& de Zeeuw (2000) proposed that the space of energy --
$z$-component of the angular momentum -- and, total angular momentum
($E$, $L_z$, $L$) would be well suited for identifying debris from
disrupted satellites. They based this conclusion on simulations of the
disruption of roughly thirty small satellites with orbits in the inner
halo, i.e. with debris contributing to the Solar neighbourhood. They
found that the particles from the satellites remained strongly
clumped, even after 10 Gyr of evolution, in the space of energy-$L_z$
and total angular momentum. Their simulations were rather simplistic,
since the satellites were evolved in a fixed potential, but
highlighted that recovery would be possible in a space of
quasi-conserved quantities. In a hierarchical Universe, the total
energy of a satellite is not a conserved quantity, since strong
fluctuations in the gravitational potential are expected to have been
important throughout the build up of galactic systems. Massive
satellites also suffer from dynamical friction, implying that the
stars lost in different passages end up having different energy
levels, an effect which is visible for example, in the work of Knebe
et al. (2005). Moreover, the total angular momentum is not a conserved
quantity either, even in the static case considered by Helmi \& de
Zeeuw, because the Galactic potential is not spherical. Hence, it is
clear that a new approach is required beyond that previously proposed.

\subsection{The APL space}
\label{sec:sims}

Let us first explore the characteristics of debris from a disrupted
satellite in the ``fixed-potential'' regime. We focus here on a
simulation that reproduces the properties of the Arcturus group and
that was presented in Navarro et al. (2004). In this simulation, we
represent the satellite by $10^5$ particles that follow a King sphere
with concentration $c = \log_{10} r_t/r_c = 1.25$ and core radius
$0.39$ kpc. Its initial 1-D velocity dispersion is $18.6$ \kms, while
its total mass is $M_{sat} = 3.75 \times 10^8 \sm$\footnote{These
numerical values differ from those quoted in Navarro et al. (2004) due
to a typographical error.}. This simulation models the self-gravity of
the satellite using a quadrupole expansion of the internal potential
(Zaritsky \& White 1988), but is otherwise run in the fixed Galactic
potential $\Phi = \Phi_{\rm halo} + \Phi_{\rm disk} + \Phi_{\rm
bulge}$, where
\begin{equation}
\label{sag_eq:halo}
\Phi_{\rm halo} = v^2_{\rm halo} \ln (1 + R^2/d^2 + z^2/d^2), 
\end{equation}
with $v_{\rm halo} = 134 $ \kms, and $d = 12$ kpc, 
\begin{equation}
\Phi_{\rm disk} = - \frac{G M_{\rm disk}}{\sqrt{R^2 + (a_{\rm d} + 
\sqrt{z^2 + b_{\rm d}^2})^2}}, 
\label{sag_eq:disk}
\end{equation}
with $M_{\rm disk} = 9.3 \times 10^{10} \sm $, $a_{\rm d}$ = 6.5 kpc
and $b_{\rm d}=0.26$ kpc, and
\begin{equation}
\Phi_{\rm bulge} = - \frac{G M_{\rm bulge}}{r + c_{\rm b}}, 
\label{sag_eq:bulge}
\end{equation}
with $M_{\rm bulge} = 3.4 \times 10^{10} \sm$ and $c_{\rm b}=0.7$~kpc.
The numerical values of the relevant parameters in these models are
chosen to provide a good fit to the rotation curve of the Milky Way.
The satellite's orbit has apocentre $\sim 9.3$ kpc, pericentre $\sim
3.1$ kpc, eccentricity $\epsilon=(r_{apo}-r_{peri})/(r_{apo}+r_{peri})
\sim 0.5$ and $L_z \sim 970$ kpc \kms.

\begin{figure}
\includegraphics[width=8.5cm,clip]{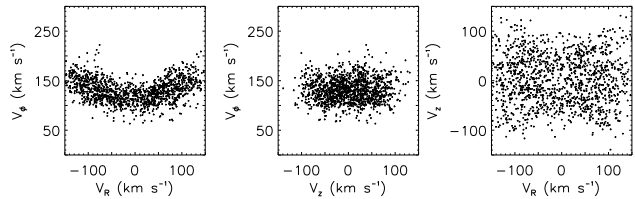}
\caption{Velocity distribution of particles from a simulated
satellite, 8 Gyr after infall. The particles are located in a sphere
centred on the ``Sun'' of 1.5 kpc radius.}
\label{fig:arct_vel}
\end{figure}

In Figure \ref{fig:arct_vel} we show the velocities of the particles
from the satellite after 8 Gyr of evolution. We focus on a sphere
centred on the ``Sun'' (at 8 kpc from the galaxy's centre) of 1.5 kpc
radius. The ``banana'' shape seen in the first panel of this figure,
defined by the $V_\phi$ and $V_R$ velocities, is due to the presence
of groups of particles with slightly different orbital
phases. Particles located at apocentre now have $V_R \sim 0$ \kms,
while those moving towards or away from their apocentres, have
positive or negative $V_R$ respectively. As we shall see later, this
very characteristic banana shape is present whenever a group of stars
have similar orbital eccentricities. Note as well from
Fig.~\ref{fig:arct_vel} the symmetry in $V_R$ and $V_z$, which
reflects the high degree of mixing of the particles. The velocity
distribution of satellite debris is notably different from that
associated to substructures produced by dynamical resonances (see
bottom panel of Fig.~\ref{fig:vel_df}). Its characteristic shape will
later on enable us to make a distinction between substructures
associated to satellite debris and those due to dynamical processes.

\begin{figure}
\includegraphics[width=8.5cm,clip]{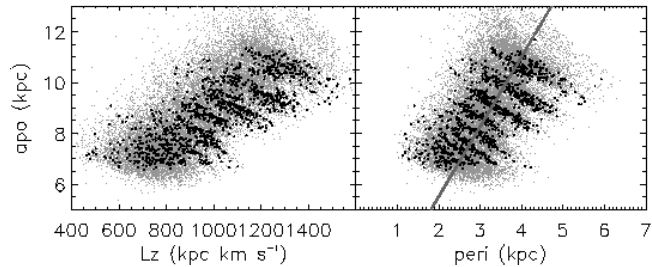}
\caption{Distribution of particles in the apocentre-pericentre-$L_z$
(APL) space for the simulation shown in Fig.~\ref{fig:arct_vel}. The
black (gray) points represent particles located in a Solar
neighbourhood volume of 1.5 kpc (5 kpc) radius centred on the Sun.}
\label{fig:arct_apl}
\end{figure}

Instead of the energy -- $L_z$ -- angular momentum space, we now focus
on that defined by orbital apocentre -- pericentre -- $L_z$: the
APL-space. After some experimentation, we found that this space
appears to be very suitable for our goals of finding accretion
relicts. Figure \ref{fig:arct_apl} shows the distribution in the
APL-space of the particles in our simulation, after 8 Gyr of
evolution. The black solid points correspond to particles located
inside a sphere of 1.5 kpc radius centred on the ``Sun'' (the same
volume as that used in Figure \ref{fig:arct_vel}), while the gray dots
are within a larger sphere of 5 kpc-radius.  The particles are not
scattered throughout the APL-space but occupy a relatively small
region. The extent of this region is determined by the initial size of
the system (since dynamical friction or other dynamical processes that
could affect the extent are not included in the simulation). In the
pericentre vs apocentre projection the particles distribute themselves
around a curve of constant eccentricity (shown as the solid line).

There is little difference between the distribution of black and grey
points in Fig.~\ref{fig:arct_apl}, which indicates that even particles
within a small volume sample the whole extent of the satellite in the
APL-space. This is because enough time has passed since infall for
these particles to phase-mix and be distributed in a number of
different streams that are crossing the particular volume under
consideration. In this case, there are of the order of ten different
streams, which can be identified as the streaks in the different
projections of the APL-space.  Particles on the same stream have
similar apocentres, pericentres and $L_z$, but these quantities are
correlated and hence not completely independent from one another.
These correlations are then reflected in the streaks (or
curves/surfaces connecting the variables) which are visible in the
various projections of the APL-space. In other words, the topology of
a stream in phase-space could, in principle, be even better
characterized using a different set of (orthogonal) variables that are
some combination of apocentre, pericentre and $L_z$.

When a satellite galaxy is accreted in a time-dependent gravitational
potential or when it suffers dynamical friction, the distribution of
its particles in the APL will be slightly different. Like in the
static case, stars lost in the same passage are expected to have more
similar orbital parameters (including pericentric and apocentric
distances) than those lost in other passages\footnote{In reality, at
each pericentric passage two groups of stars are released, defining a
trailing and a leading stream. The properties of the stars in each of
these streams are expected to be similar, while the energies of the
trailing and leading streams should be offset (Johnston 1998; Mayer et
al. 2002).}. Changes in the Galactic potential are likely to occur on
timescales longer than one orbital period of the satellite, implying
that those stars lost in the same passage will remain clumped together
and presumably perturbed onto a different orbital path (with larger
binding energy if the galaxy has, for example, gained a large amount
of mass in the inner regions during a major merger). A similar
argument holds if dynamical friction is important. Massive satellites
will deposit stars in different energy levels, according to the
time/passage when the stars were released. So, to first order, one
expects stars to generally cluster in the APL space and along a line
of constant eccentricity also in a time-dependent situation.

\subsection{Galaxy formation simulations}
\label{sec:sims_cosm}

It is important to confirm our conclusions about the evolution of
debris in time-independent gravitational potentials.  We need to
establish whether the APL space is a conserved quantities space also
under the conditions found in a hierarchical universe.

Hence, it is necessary to turn to simulations of the formation of
galaxy disks which take into consideration all possible dynamical
effects, including the evolution of the Galactic potential and
dynamical friction, collisions between satellites, etc. Abadi et
al. (2003a) present one such high resolution simulation that models
the formation of a disk galaxy, albeit of an earlier type than the
Milky Way.  Nevertheless, we expect that the gross characteristics of
the dynamical evolution of substructure will not depend strongly on
the exact properties of the main galaxy (provided the simulation
represents a disk and not, e.g. a highly triaxial system).  An
important ingredient of this simulation is the self-consistent
modeling of star formation, which allows the distinction between star
and dark-matter particles. This is advantageous for our purposes since
we expect the identification of stellar substructures to be easier
than for their dark-matter counterparts. This is because the
dark-matter occupies a larger region of phase-space than the stellar
components of galaxies (which are typically more concentrated to the
centres of the halos in which they are embedded). Hence, stellar
substructures should have a higher density contrast and be less
convoluted (spatially) than the dark-matter. All this implies that the
chance of superposition of lumps in the APL-space ought to be smaller
for stars than for dark-matter.
\begin{figure}
\includegraphics[width=8.5cm,clip]{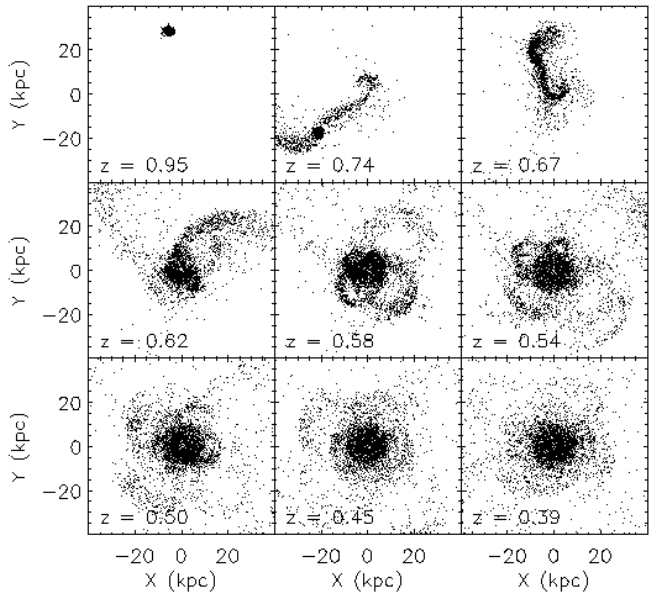}
\caption{Face on view of the spatial distribution of the
star-particles from a satellite identified in a high resolution
cosmological simulation. The satellite starts to merge with the host
galaxy at $z = 1$, and after a few orbits it is completely disrupted.}
\label{fig:xy_sat}
\end{figure}
\begin{figure}
\includegraphics[width=8.5cm,clip]{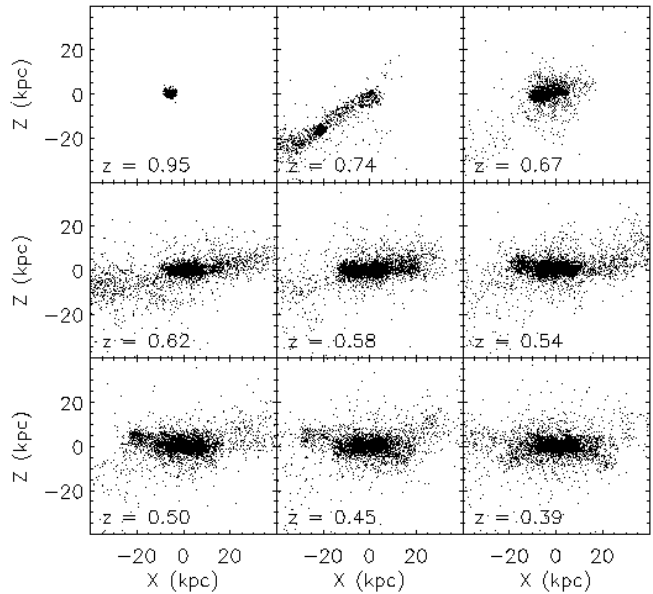}
\caption{The same satellite as Fig.~\ref{fig:xy_sat} but now the
distribution of star-particles is shown in an edge-on view. Most of
the debris of the satellite is deposited in a thick disk-like
configuration.}
\label{fig:xz_sat}
\end{figure}

We have selected ten satellites in the simulation by Abadi et
al.~(2003a), with masses in the range $6 \times 10^9 \sm$ to $2.7
\times 10^{10} \sm$, which merge with the central galaxy in the
redshift range 3.4 to 1. Figures~\ref{fig:xy_sat} and \ref{fig:xz_sat}
show the evolution of one of these satellites. Its total mass at the
time of infall (at redshift $z=1.05$) was $1.1 \times 10^{10} \sm$,
while its stellar mass (represented by $\sim 5000$ particles) was $4.5
\times 10^{9} \sm$. Efficient dynamical friction brings it to the
inner galaxy, and so this satellite contributes with star-particles to
the Solar circle. A fair fraction of this debris is deposited on
eccentric orbits in the galactic disk (Abadi et al. 2003b).

\begin{figure}
\includegraphics[height=9.3cm,clip,angle=90]{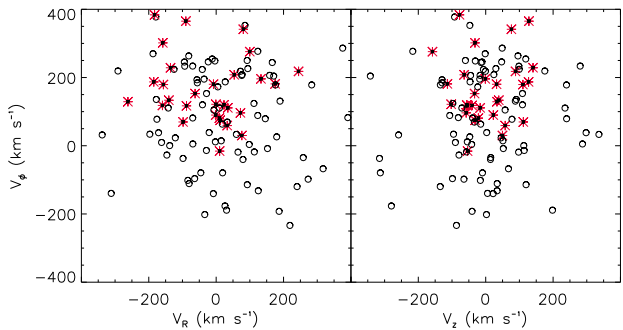}
\vspace*{-1.4cm}
\caption{Velocity distribution of star-particles located in a box
centred on the equivalent of the Solar vicinity. We show the particles
originating in ten satellites identified in the cosmological
simulation (open circles), and highlight (asterisks) the contribution
of the satellite shown in Figs.~\ref{fig:xy_sat} and \ref{fig:xz_sat}.}
\label{fig:box_vel}
\end{figure}

 In Figure~\ref{fig:box_vel} we show the present-day velocities of the
star-particles originating in those ten satellites, for a 4 kpc box
centred on the Solar circle. We have highlighted the contribution of
the satellite shown in Figs.~\ref{fig:xy_sat} and
\ref{fig:xz_sat}. There are some hints of moving groups (around
$V_\phi \sim 80$ \kms, $V_\phi \sim 120$ \kms and $V_\phi \sim 200 $
\kms), but clearly the number of particles is too small to fully
resolve the velocity distribution into substructures.

We now proceed to determine the distribution in the APL space for the
star-particles in our simulation. We compute the apocentric and
pericentric distances through orbital integration in the Galactic
potential introduced in Sec.~\ref{sec:sims}.  Note that this potential
does not constitute a very good model of our simulated galaxy, which
is of earlier-type (Sa) than our Galaxy. However, this is done
intentionally since observers will only approximately know the form of
the underlying potential.

\begin{figure}
\hspace*{-0.7cm}
\includegraphics[height=9.3cm,angle=90]{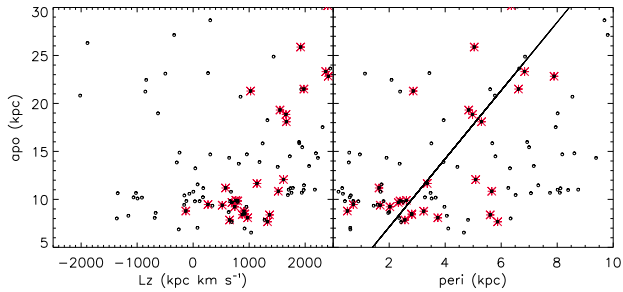}
\caption{APL-space distribution of star-particles located in a box
centred on the Solar circle. We show the particles originating in ten
satellites identified in the cosmological simulation (open circles),
and highlight (asterisks) the contribution of the satellite shown in
figures \ref{fig:xy_sat} -- \ref{fig:box_vel}. Notice that the
distribution is much lumpier than the velocity distribution shown in
Fig.~\ref{fig:box_vel}. The horizontal streaks visible in the apo-peri
plot, particularly for the asterisks, correspond to groups of
star-particles which have, presumably, become unbound around the same
time. Note as well the diagonal pattern defined by the asterisks in
both projections of the APL-space. This pattern is the same as that
shown in Fig.~\ref{fig:arct_apl} for the satellite evolving in a
static Galactic potential. The pattern shows that the debris is
distributed along a line of (quasi)-constant eccentricity as indicated
by the solid line in the panel on the right.}
\label{fig:box_all_apl}
\end{figure}

In Figure \ref{fig:box_all_apl} we show the APL-space distribution of
star-particles located in the same 4 kpc box we showed before. The
asterisks in this Figure correspond to the star-particles originating
in the satellite shown in Figs.~\ref{fig:xy_sat}, \ref{fig:xz_sat} and
\ref{fig:box_vel}.  The distribution is clearly lumpy, in stark
contrast to the velocity distribution shown in Fig.~\ref{fig:box_vel},
providing support for our hypothesis that the APL is a space of
quasi-conserved quantities. Note that this satellite has produced
several lumps in the APL space: the ``horizontal stripes'' in the
peri-apo presumably correspond to star-particles released in a given
perigalactic passage.  Such features are common to all satellites, as
evidenced by the distribution of open circles in
Fig.~\ref{fig:box_all_apl} and which corresponds to the ten satellites
identified in the simulations that contribute to this volume.

\section{Analysis of the N04 sample}
\label{sec:analysis}

\subsection{Distribution of the data in the APL space}
\label{sec:apl}

The analysis performed on the galaxy formation simulations gives us
confidence that even in the fully hierarchical regime, the
substructure associated to past mergers remains relatively intact in
phase-space. The signatures from (or left by) accreted satellites
remain coherent in the APL-space.

\begin{figure*}
\includegraphics[height=0.33\textheight,clip]{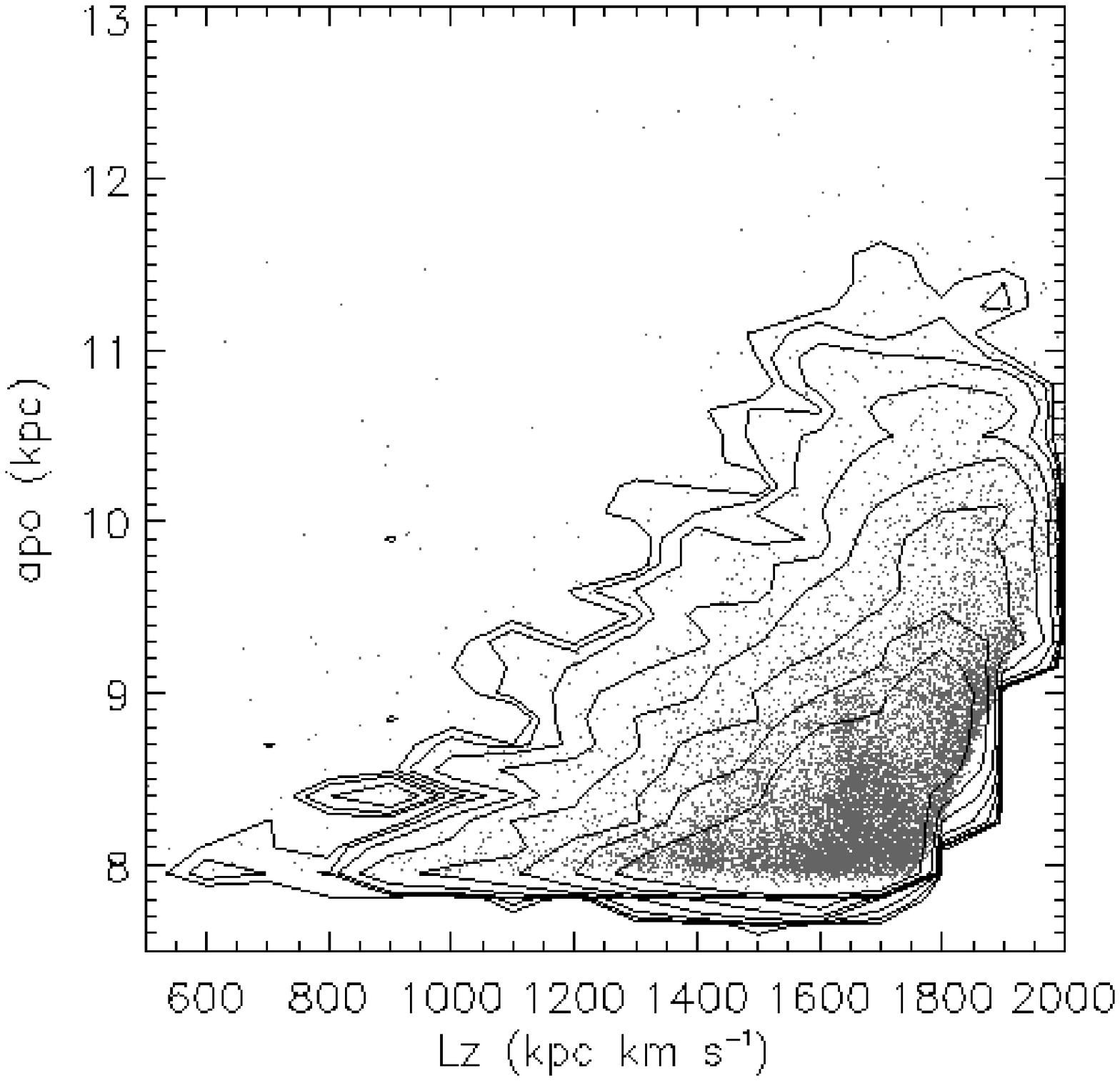}
\includegraphics[height=0.33\textheight,clip]{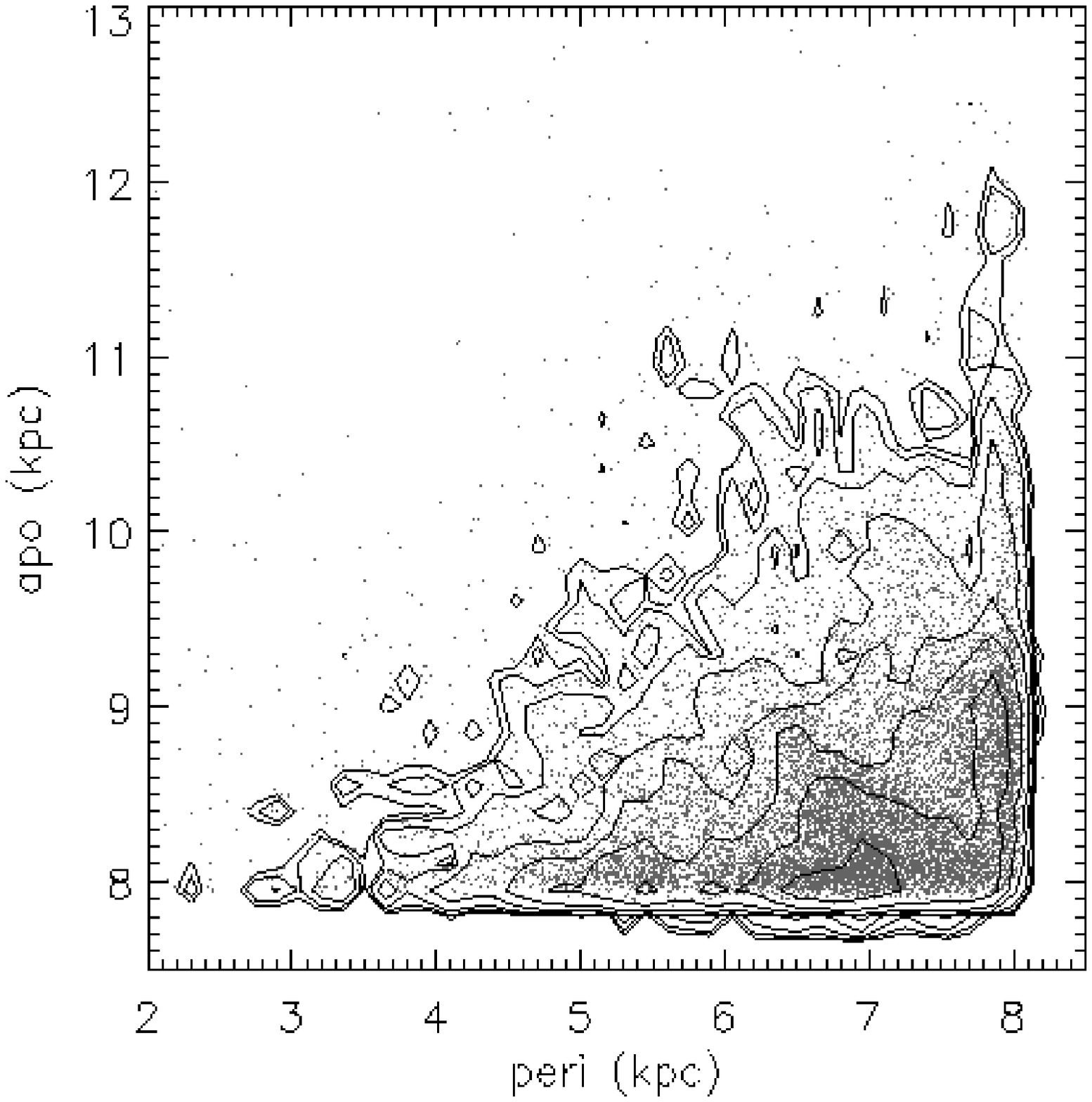}
\caption{Distribution of the stars in the N04 sample in the APL
space. The contours help visualize the very lumpy nature of the
sample. }
\label{fig:data_apl_co}
\end{figure*}

Figure~\ref{fig:data_apl_co} shows the distribution in the APL space
of the stars in the N04 sample. The largest uncertainties in the
location of a star in this space are not due to observational errors
(which are typically less than 1.5 \kms in the velocity), but in the
limited knowledge of the form of the Galactic potential used to
determine the orbital parameters (apocentre and pericentre).  However,
we do not expect this uncertainty to strongly affect the distribution
of points in the APL-space. The volume probed by the N04 sample is so
small that the Galactic potential is close to constant inside this
region.  This implies that the energy, and hence the orbital
parameters or the location of a star in the APL-space, are determined
mostly by its kinematics, rather than by its spatial location (or the
Galactic potential). Changes in the Galactic potential produce only
small variations in the orbital parameters. For example, in the case
of the potential proposed by Flynn et al. (1996), we find a typical
change of 1-2\% in apocentre and pericentre.

Small moving groups, such as the Hyades open cluster, define very
tight structures in phase-space, and this is also reflected in their
characteristic sizes in the APL-space. For example, the stars in the
Hyades open cluster have apocentre $\sim 8.52$ kpc, pericentre $\sim
6.85$ kpc and $L_z \sim 1659$ kpc \kms, while the dispersions around
these mean values are $\sigma_{\rm apo} \sim 0.05$ kpc, $\sigma_{peri}
\sim 0.07$ kpc and $\sigma_{L_z} \sim 11$ kpc \kms. On the contrary,
substructures due to past mergers will generally be more extended in
APL-space, as evidenced in Figs.~\ref{fig:arct_apl} and
\ref{fig:box_all_apl}, where typically $\Delta_{apo,\,peri} \sim 0.5 -
1$ kpc and $\Delta L_z \sim 200 - 400$ kpc \kms.

It is clear from Figure~\ref{fig:data_apl_co} that the distribution of
stars in the APL-space is rather lumpy, as shown by the iso-density
contours.  A large fraction of the substructures are not only found in
the ``tails'' of the distribution of data points, but also right where
the density of points is highest. The most prominent of these
structures are actually due to the Hyades-Pleiades, Sirius and
Hercules superclusters that were identified in
Fig.~\ref{fig:vel_df}. These occupy a region of the APL-space
dominated by disk-like and low eccentricity orbits, with $\epsilon \le
0.2$.

There are two issues that need to be addressed before accepting the
hypothesis that the substructures observed in
Fig.~\ref{fig:data_apl_co}, and which are not associated to open
clusters or superclusters, are due to past mergers. The first issue is
to establish how many of those substructures are simply statistical
fluctuations of otherwise smooth distributions.  This is addressed in
Sections \ref{sec:mc} and \ref{sec:stats} through Monte Carlo
simulations. A second issue concerns the interpretation of these
structures as satellite debris. We will address this in Sections
\ref{sec:gral} and \ref{sec:satellites}, where we look in detail into
the properties of the different structures discovered.

\begin{figure*}
\includegraphics[height=0.33\textheight,clip]{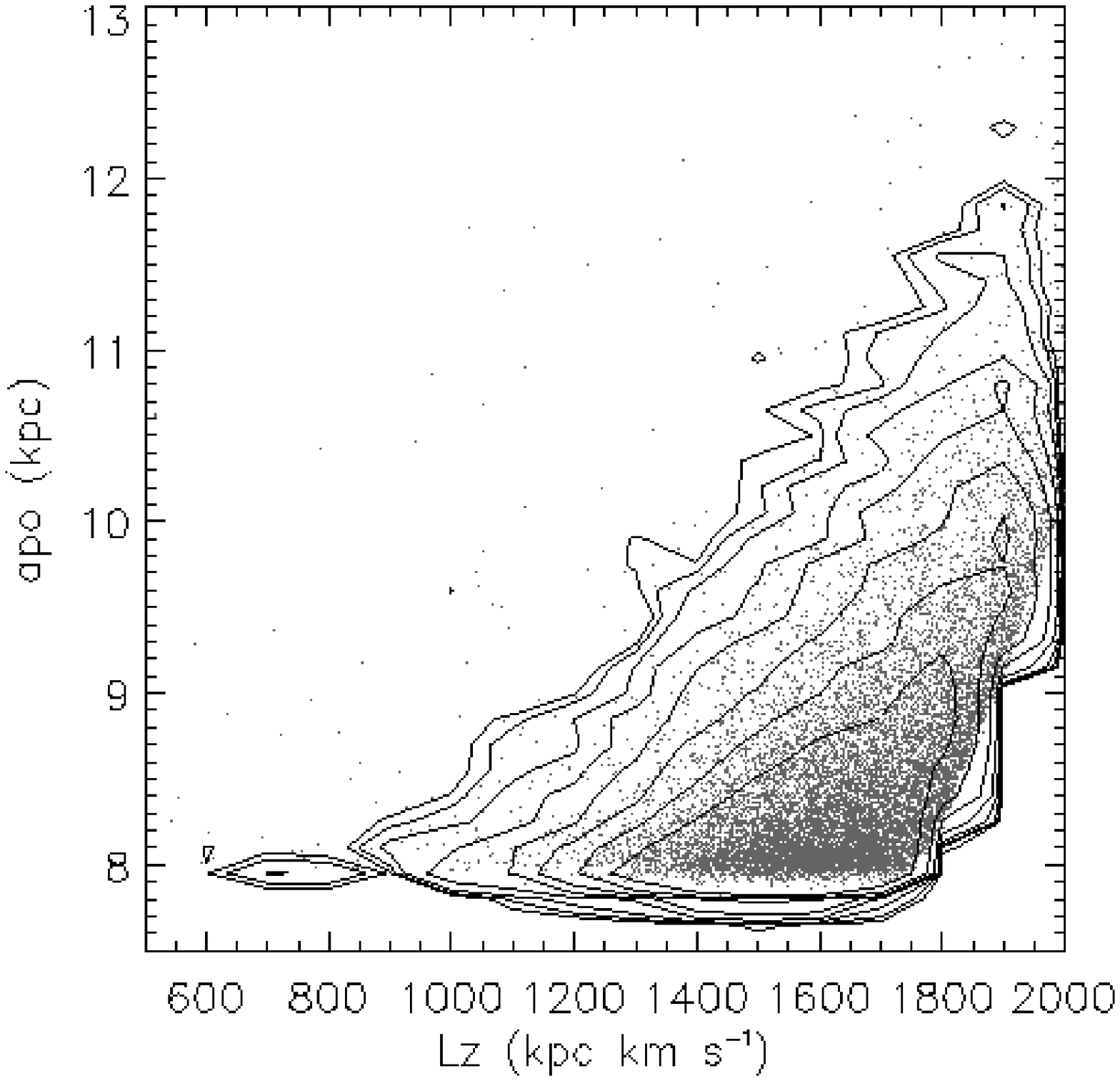}
\includegraphics[height=0.33\textheight,clip]{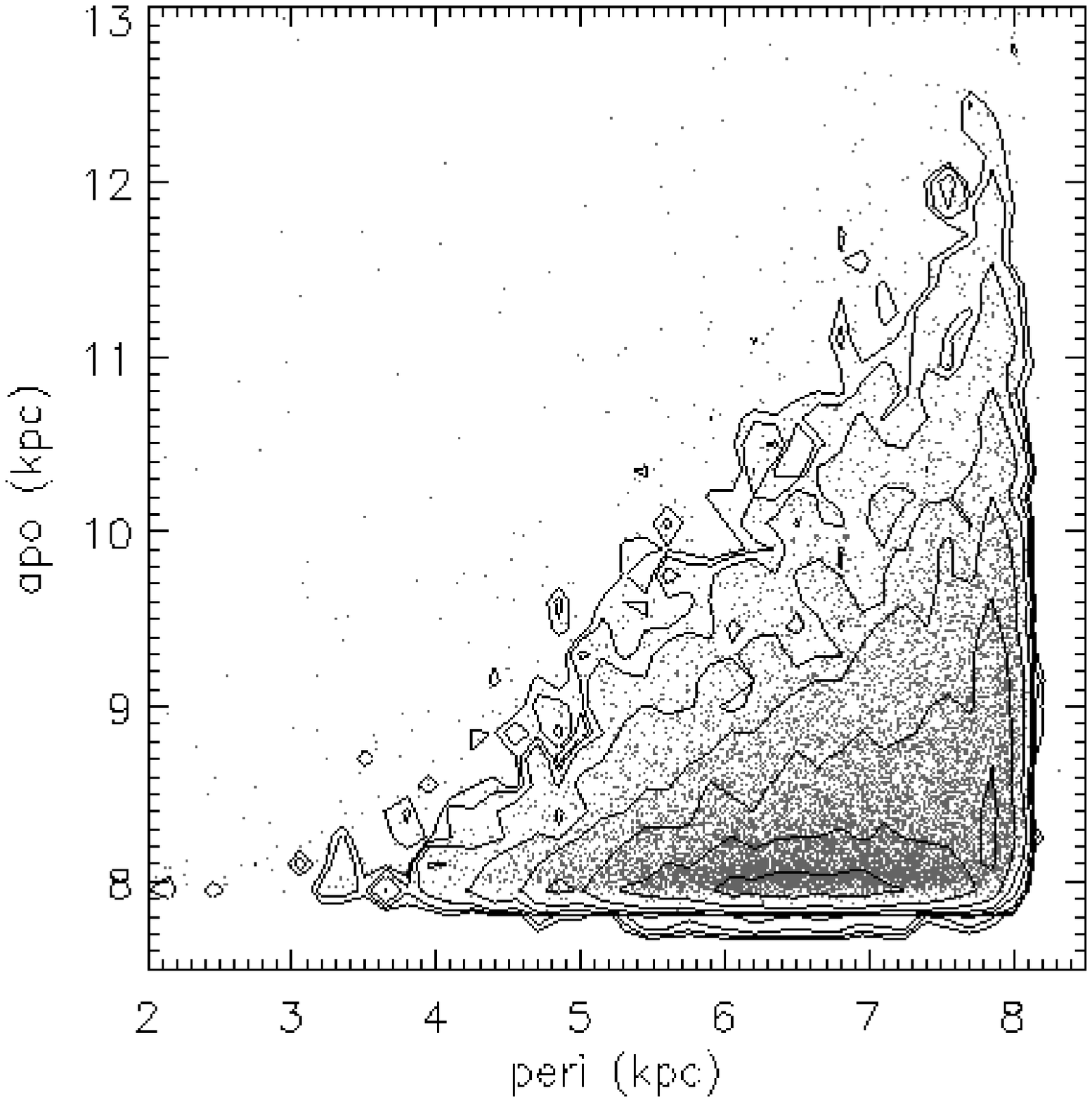}
\caption{Distribution of points from one of our Monte Carlo
simulations of a smooth Galaxy. The contours help to visualize the
degree of lumpiness of the distribution in this space. The
distribution of points with disk-like kinematics is much smoother than
that of the N04 sample shown in Fig.~\ref{fig:data_apl_co}. The contour
levels are the same as in Fig.~\ref{fig:data_apl_co}.}
\label{fig:MC_apl_co}
\end{figure*}

\subsection{Monte Carlo simulations of a smooth galaxy}
\label{sec:mc}

It is extremely important to understand the expected characteristics
of a smooth Galaxy for a dataset like the N04 sample. This is crucial
for a proper statistical assessment of the substructure observed in
Fig.~\ref{fig:data_apl_co}. To produce a smooth model of the Milky
Way, one may proceed by generating individual structural components:
thin and thick disk, halo and bulge components each with their
characteristic spatial, kinematic and metallicity distribution (Robin
et al. 2003). However, we prefer to follow a simpler approach that
consists of combining the observed spatial distribution of the stars
in the sample, with random velocities that reproduce the behaviour of
the velocity ellipsoid with metallicity shown in Figure \ref{fig:mc}.

As previously discussed, the spatial distribution of the stars in the
sample is relatively smooth, and hence we can use it directly as input
for our Monte Carlo models of the Galaxy.  On the contrary, the
observed kinematic distribution is rather lumpy (as shown in
Fig.\ref{fig:vel_df}).  Therefore, we only need to replace this
distribution by a multivariate Gaussian to obtain a smooth model of
the Galaxy. It is worth noting that we do not expect any correlations
between the spatial and kinematic distribution of the stars, because
the volume probed by these observations is sufficiently small that the
various Galactic components cannot be distinguished solely on the
basis of the spatial distribution of the stars in the sample.

The velocity distribution in our Monte Carlo simulations is
represented by a set of multivariate Gaussians whose characteristic
parameters are allowed to vary with metallicity in the manner given by
the data itself (see Fig.~\ref{fig:mc}). The orientation of the
principal axes of the velocity ellipsoid do not perfectly coincide
with the $U,V,W$ directions, and vary slightly in each metallicity
bin. Our 1000 Monte Carlo simulations have the same number of stars in
each metallicity bin as the original data, as well as the same spatial
distribution. Thus each ``Monte Carlo star'' is given a random
velocity as described above, and assigned spatial coordinates from the
original dataset.

\subsection{Statistical analysis of the sample}
\label{sec:stats}

We are now ready to compare the distribution of the observed data
points and the smooth Galaxy model in the APL-space. We compute the
orbital properties in the Galactic potential characterized by
Eqs.(\ref{sag_eq:halo} -- \ref{sag_eq:bulge}), for each of the 13092
points in our 1000 Monte Carlo samples\footnote{The number of stars
used to define the Monte Carlo simulations differs from the number of
stars with full phase-space information in the N04 sample (13092
instead of 13240), because there are 148 stars that do not have a
metallicity estimate.}.  Figure~\ref{fig:MC_apl_co} shows an example
of the distribution of points in the case of the smooth Galaxy
model. The overall characteristics (extent, high concentration of
points on disk-like orbits) of the distribution are in good agreement
with the observations (see Fig.~\ref{fig:data_apl_co}), and support
the assumptions made to generate a smooth model of the
Galaxy. Eye-ball inspection shows that the distribution is not
completely featureless (particularly in the apocentre vs pericentre
projection). A direct comparison to Fig.~\ref{fig:data_apl_co}
suggests, however, that the N04 sample contains far more substructure
(also in the regions that would be dominated by the young-disk stars,
i.e. pericentre $\sim 6 - 8$ kpc and apocentre $\sim 8$ kpc).

We now wish to establish which regions of this 3-dimensional space
show a statistically significant excess of stars compared to a smooth
Galaxy. For this quantitative analysis, we make a partition of the APL
space into cells, in which we perform number counts. We test three
different partitions in each dimension: $(45, 135, 225) \times (19,
52, 85) \times (6, 18, 30)$ in the apocentre, pericentre and
$L_z$-directions, respectively. These correspond to a resolution
(cell-size) of (0.5, 0.17, 0.1)[kpc] $\times$ (0.45, 0.16, 0.1) [kpc]
$\times$ (500, 166.7, 100) [kpc \kms], respectively.  These cell-sizes
are comparable in extent to the substructures visible in the N-body
simulations shown in Figs.~\ref{fig:arct_apl} and
\ref{fig:box_all_apl}; they are typically much larger than expected
for open clusters although comparable to the extent of the
superclusters, as discussed in Sec.~\ref{sec:apl}.  For each cell $j$
we compute the observed number of stars $N_j^{obs}$, and the average
number of points from the Monte Carlo realizations $\langle N_j
\rangle$ and the dispersion around this average, $\sigma_j$. To
establish which cells have a statistically significant excess of
stars, we define $s_j$ and $p_j$ as
\begin{equation}
s_j = \frac{N_j^{obs} - \langle N_j \rangle}{\sqrt{N_j^{obs}+ \sigma_j^2}}
\end{equation}
which measures the excess of stars in each cell, weighted by the average
expected ``dispersion'' in the cell, and
\begin{equation}
p_j = \frac{e^{-\lambda_j} \lambda_j^{x_j}}{\Gamma(1 + x_j)}
\end{equation}
where $\lambda_j = \langle N_j \rangle$ and $ x_j =
N_j^{obs}$. Therefore $p_j$ measures the probability of observing
$x_j$ stars in cell $j$ if the counts follow a Poisson distribution,
whose characteristics are determined by the distribution of points
from the Monte Carlo simulations. When the number of stars in a given
cell is large, one may assume that the Gaussian limit has been reached
and then $s_j$ is the right statistic to use. In that case, we
consider cells for which $s_j \ge 3$ to show a statistically
significant overdensity. However, when the number of stars in a cell
is not very large (typically less than 20), it is best to use the
second statistic, and to identify significantly overdense cells as
those for which $p_j \le 0.01$ and $s_j > 0$.

\begin{figure}
\includegraphics[width=8.5cm]{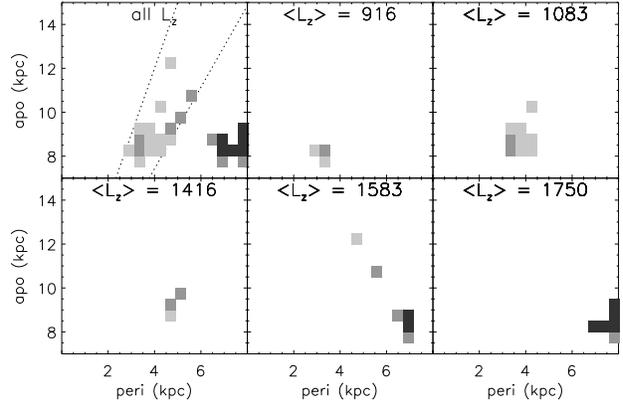}
\vspace*{-3cm}
\caption{Distribution of overdense cells in the APL space. In the
first panel we show a projection of the cells onto
apocentre-pericentre space, while the rest of the panels correspond to
slices in $L_z$, i.e. a narrow range of $L_z$ values, and hence help
visualize the 3D nature of the APL space. The colour-coding indicates
the statistical significance in steps of 10: black corresponds to $p <
10^{-5}$, dark grey to $10^{-5} \le p < 10^{-4}$, grey to $10^{-4} \le
p < 10^{-3}$ and light grey to $10^{-3} \le p < 10^{-2}$. Notice the
diagonal patterns of overdensities that extend from apocentre $\sim 8
- 9$ kpc, pericentre $3.5 - 4$ kpc and which also have increasing
$L_z$. They are very reminiscent of the structures produced by
disrupted satellites and shown in Figs.~\ref{fig:arct_apl} and
\ref{fig:box_all_apl}. The overdense cells are located in a segment of
the APL space delimited by eccentricity $\epsilon \sim 0.3 - 0.5$,
as indicated by the dotted lines in the first panel.}
\label{fig:cells}
\end{figure}

In Figure \ref{fig:cells} we have plotted the location of the cells
that show a significant overdensity according to the above
criteria. Because of the 3-d nature of the APL, we show slices of the
apocentre vs pericentre space.  This figure corresponds to a partition
of $45 \times 19 \times 18$ cells, or equivalently, to cell sizes of
0.5 kpc $\times$ 0.45 kpc $\times$ 166.7 kpc \kms. The colour-coding
indicates how statistically significant the overdensity is: from
$10^{-3} \le p < 10^{-2}$ (light grey) to $p < 10^{-5}$ (black).  The
regions identified also appear as significant overdensities for other
choices of the partition. We are therefore confident that these
results do not depend strongly on the cell-size.

The cells that show the most significant overdensities are located in
a region of the APL space which is largely dominated by thin disk
stars (bottom central and right panels), and which in fact coincides
with the location of the superclusters as defined by Famaey et al.\
(2005). The presence of overdensities in this region is likely the
result of dynamical perturbations induced by asymmetries in the
Galactic potential, such as the spiral arms and the Galactic bar, as
shown by Dehnen (1998) and De Simone et al. (2004). Clearly, a
Gaussian distribution does not provide a good representation of the
kinematics of thin disk stars (of young to intermediate age), and it
will be interesting to try and reproduce the features observed here
with very detailed dynamical modelling and specify from more general
principles the distribution function and its evolution in phase-space
(see Quillen \& Minchev 2005, for a recent attempt).

Another intriguing feature of Fig.~\ref{fig:cells} are the
overdensities located on a diagonal pattern in the apocentre vs
pericentre projection, and which have progressively larger $L_z$.
This pattern is very reminiscent of that produced by satellite debris,
as shown in Figs.~\ref{fig:arct_apl} and \ref{fig:box_all_apl}. It is
therefore natural to believe that these substructures are indeed the
remains of disrupted satellites.

As discussed in Secs.~\ref{sec:sims} and \ref{sec:sims_cosm}, such
overdensities fall close to a line of constant eccentricity, which is
reflected in the diagonal pattern observed in the projections of the
APL-space.  In the observations, the overdense cells are limited by
eccentricities $\epsilon_{min} \sim 0.3$ and $\epsilon_{max} \sim 0.5$
(the dotted lines in the first panel of Fig.~\ref{fig:cells}).

\begin{figure}
\hspace*{-0.5cm}
\includegraphics[width=9cm]{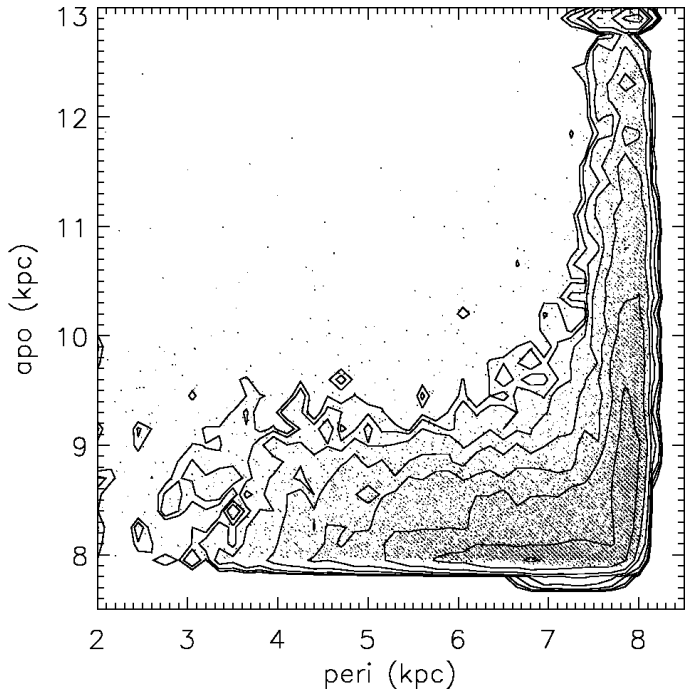}
\caption{Distribution of points in the apocentre-pericentre projection
of the APL space for one realization of the Besancon model of the
Galaxy. By comparing to the panel on the right in
Fig.~\ref{fig:data_apl_co} we note that this distribution is
significantly different than that found for the N04 sample. }
\label{fig:besancon}
\end{figure}

One may question whether these overdensities are real, or simply the
result of a poor choice of the comparison (smooth) model of the
Galaxy. In an attempt to understand this, we turn to the Besancon
model of the Galaxy. This is a standard model of the Galaxy, with thin
and thick disk, as well as bulge and halo components (Robin et
al. 2003, 2004). We have used the machinery provided on the Besancon
website ({\sf http://bison.obs-besancon.fr/modele/}) to generate
samples of ``stars'' with the same magnitude and colour limits as the
N04 sample. We then calculate their orbital parameters in the Galactic
potential given by Eqs.~\ref{sag_eq:halo} - \ref{sag_eq:bulge}, and
determine their distribution in the APL-space. Through this exercise,
we find that the Besancon Galaxy model has an even smaller number of
``stars'' in the region occupied by the overdensities, as shown in
Fig.~\ref{fig:besancon}. Therefore, even for two rather different ways
of making smooth Galaxy models, we obtain significant overdensities
along a constant eccentricity segment in the APL-space. This suggests
that our interpretation of the data is statistically robust and may be
the only viable one.

\subsection{Focus on the overdensities}
\label{sec:gral}

\subsubsection{Metallicity distribution}
\begin{figure}
\vspace*{-0.5cm}
\hspace*{-0.3cm}
\includegraphics[width=9cm]{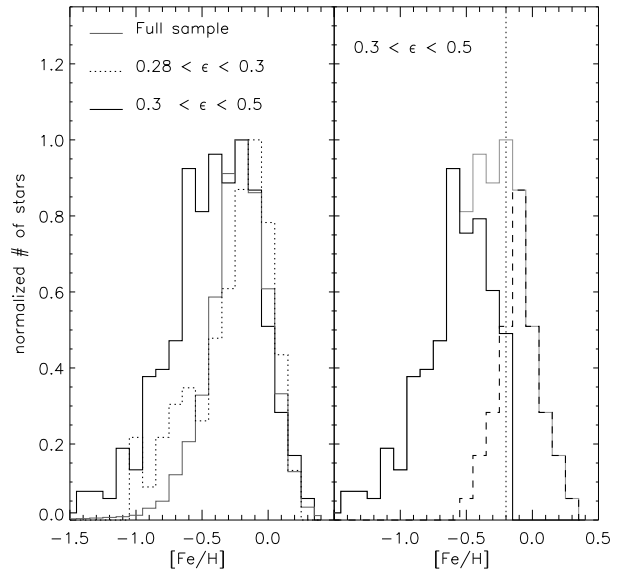}
\caption{{\it Left: }[Fe/H] distribution for the full N04 sample
(grey), for the stars located in the segment of eccentricity defined
by the overdense cells in Fig.~\ref{fig:cells} (black), and for a
comparison set (dotted). Note the distinctly different metallicity
distribution of the stars located in the overdensities in comparison
to the full sample, and in particular to the set that has just
slightly different orbital characteristics. {\it Right: } The solid
black histogram shows the metallicity distribution of the stars
located in the segment $\epsilon \in [0.3,0.5)$, after subtracting the
contribution of the thin disk (dashed). The vertical dotted line
corresponding to [Fe/H]$=-0.2$~dex, serves to indicate that to its
left, the contamination by thin disk stars is small.}
\label{fig:feh}
\end{figure}

We now proceed to study in more detail the properties of the stars
located in this segment of eccentricities. The left panel of
Figure~\ref{fig:feh} shows the metallicity distribution of these stars
(black histogram). For comparison we also plot the distribution of a
control set of rather similar kinematics ($0.28 < \epsilon < 0.3$,
dotted histogram), and the metallicities of the full sample (grey
solid histogram).  Note that the metallicity distribution of the full
sample is rather narrowly concentrated around a peak value of
[Fe/H]$\sim -0.2$ dex, which is close to the mean metallicity of the
thin disk in this sample, i.e. $\langle$[Fe/H]$\rangle_{disk}\sim
-0.13$ dex (derived from stars with $|V_z| \le 30$ \kms and
$\epsilon \le 0.1$). The metallicity distribution of stars in the
selected segment of eccentricities $\epsilon \in [0.3,0.5)$ is much
flatter and broader, and therefore significantly different. In fact,
the probability that the selected set has been randomly drawn from the
full sample, as measured by a KS test (Press et al. 1988) is less than
$10^{-8}$.

The metallicity distribution of the control set is also strikingly
different from that of the selected set, despite their small
difference in orbital eccentricities.  It resembles more closely the
distribution of the full sample, although it has a more pronounced
tail towards lower metallicities.  The KS test applied to the control
and selected sets gives a probability that the two have the same
parent metallicity distribution of approximately $10^{-4}$, while the
probability for the control and the full samples is 3\%. Clearly,
there is a sharp transition in the properties of stars above and below
$\epsilon \sim 0.3$, not just in terms of the degree of dynamical
substructure but also in their metallicity distribution. The stars
located in the region defined by the overdense cells with $0.3 \le
\epsilon < 0.5$ appear to constitute a separate class.

 Even in this relatively high eccentricity range, contamination by
thin disk stars is likely to be important. In the panel on the right
of Figure~\ref{fig:feh} we plot again the metallicity distribution of
stars in the selected segment of eccentricities (grey histogram). It
is safe to infer that in this set, all stars with
[Fe/H]$>\langle$[Fe/H]$\rangle_{disk}$ belong to the thin disk. By
assuming the stars are symmetrically distributed with respect to
$\langle$[Fe/H]$\rangle_{disk}$, we can actually subtract their
contribution (dashed histogram). Therefore, we obtain a ``cleaner''
metallicity distribution of the stars in the selected segment of
eccentricities (shown as the solid black histogram). This distribution
shows a prominent peak at [Fe/H]~$\sim -0.6$ and a broad ``bump''
around [Fe/H]~$\sim -0.8$. For the sake of simplicity, in what follows
we will restrict further analysis to stars in the selected segment of
eccentricities with [Fe/H]~$< -0.2$ dex (i.e. to the left of the
vertical dotted line in Figure~\ref{fig:feh}).

\subsubsection{HR diagrams}
\begin{figure}
\hspace*{-0.5cm}
\includegraphics[width=8.7cm]{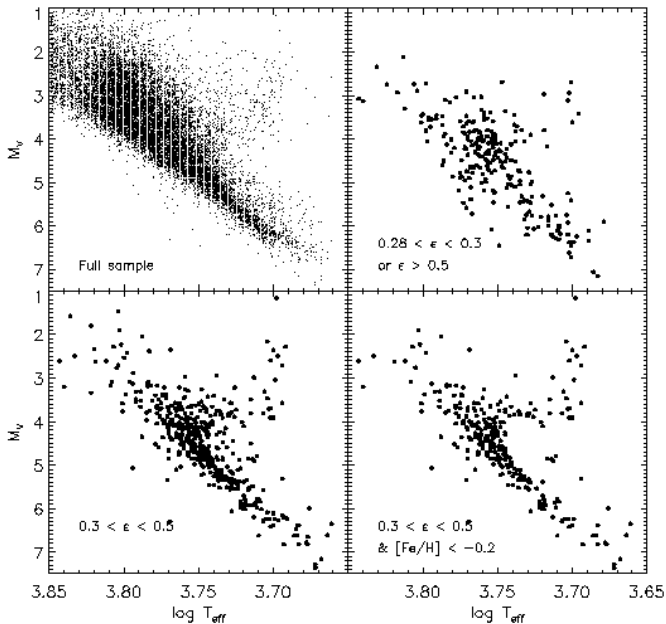}
\caption{Hertzsprung-Russell diagram of the full N04 sample (top
left); of a comparison set with lower/higher eccentricities than the
overdense segment (top right); of the set of stars located in the
overdense segment (bottom left); and a subset of the latter with lower
metallicity. The bottom panels show the well populated red giant
branch and the dearth of luminous stars above $M_V \sim 3.8$, which
suggests the presence of a turn-off point around that luminosity.
Furthermore, notice how different the HR diagrams on the top
and bottom rows are.}
\label{fig:cmd}
\end{figure}

Further insight into the properties of the stars located in the
overdense segment of eccentricities can be obtained from the
Hertzsprung-Russell diagram. Effective temperatures have been
determined from the Str\"omgren colours using the indices and
calibrations of Alonso et al. (1996), while the absolute magnitudes
are known thanks to the distance information available for the stars
in the sample.

The top left panel of Figure \ref{fig:cmd} shows the HR diagram of the
stars in the N04 sample, while that on the top right corresponds to a
comparison set with eccentricities which are slightly different from
those found in the overdense cells. The panel on the lower left shows
the HR diagram for all stars in the overdense segment of
eccentricities, i.e. with $0.3 \le \epsilon < 0.5$, while that on the
right corresponds to the subset of these stars that have [Fe/H]
$<-0.2$ dex (minimizing the contamination by the thin disk). The two
most striking features in the lower HR diagrams are the presence of a
well-defined red giant branch (perhaps more than one), and a possible
corresponding turn-off around $M_V \sim 3.8$, which is visible thanks
to the apparent dearth of stars with higher luminosities (younger
ages). The elimination of stars with high metallicity further enhances
the coherent features in the HR diagram of the overdense eccentricity
segment, and provides more confidence that we have identified the
debris of one or more disrupted galaxies. The stars in the overdense
regions have a strikingly different age distribution than their parent
sample (and than the control set shown in the top right panel).

\subsection{Characteristics of the satellites}
\label{sec:satellites}
\begin{figure}
\includegraphics[width=8cm]{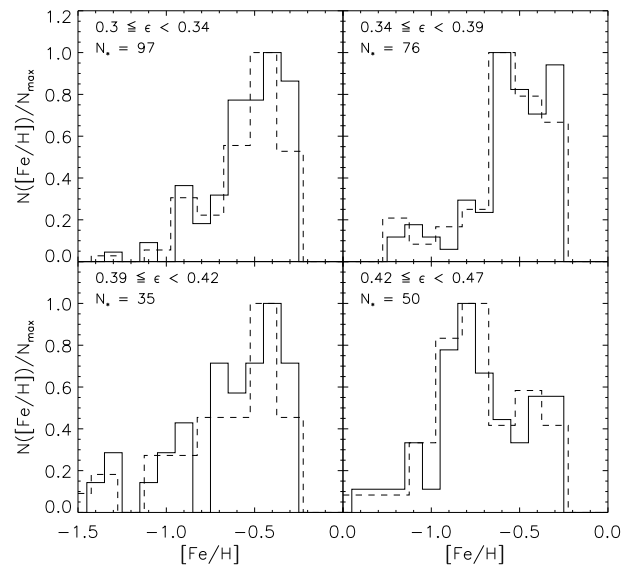}
\caption{Distribution of metallicities for the stars located in the
overdense segment of eccentricity identified in
Sec.~\ref{sec:stats}. Only those stars more metal-poor than
[Fe/H]~$=-0.2$ dex are shown. The different panels correspond to cuts
in eccentricity, while the solid and dashed histograms are for
metallicity bins of 0.1 and 0.15 dex, respectively. Note the
distinctly different metallicity distribution of the stars with lowest
(top left) and the highest (bottom right) eccentricities.}
\label{fig:feh_ecc}
\end{figure}

\subsubsection{How many?}
\begin{figure*}
\vspace*{1.5cm}
\includegraphics[width=17.5cm]{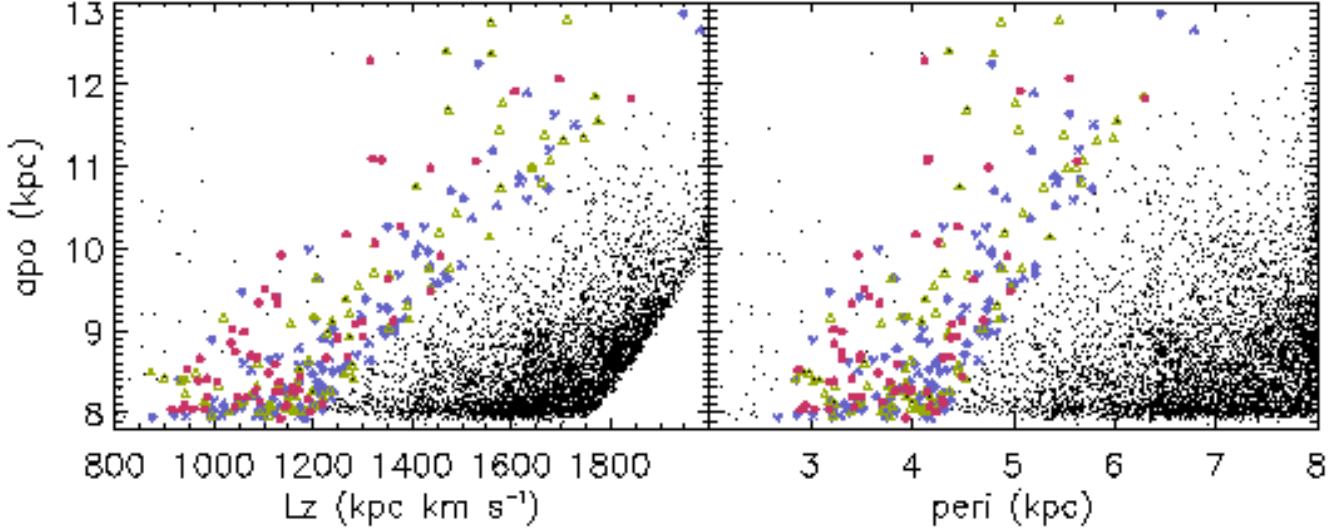}
\vspace*{0.5cm}
\caption{Identification of stars associated with the overdense region of
constant eccentricity in the APL space. The (blue) asterisks
correspond to Group 1 (120 stars), with [Fe/H]~$\in [-0.45,-0.2)$ dex;
the (green) triangles to Group 2 (86 stars): [Fe/H]~$\in [-0.7,-0.45)$
dex; and the solid dots to Group 3 (68 stars): [Fe/H]~$\in
[-1.5,-0.7)$ dex. Note that the various groups are distributed
slightly differently in the APL-space. The black small dots correspond
to all stars in the N04 sample with [Fe/H]$<-0.2$~dex.}
\label{fig:apl_groups}
\end{figure*}

One of the most interesting questions in the context of the formation
of the Galaxy in a hierarchical Universe, is how many progenitors has
it had and what were their characteristics? Therefore, it is very
important to gauge the number of satellites that have given rise to
the structures with eccentricity $0.3 \le \epsilon < 0.5$. It is clear
from the metallicity distribution shown in Fig.~\ref{fig:feh} that the
satellites that have contributed to this region of phase-space, must
have been relatively massive to be able to contribute a large number
of stars with [Fe/H]~$\sim -0.5$ dex. The extent of the overdensities
in the APL-space are similar to that seen in the simulation discussed
in Sec.~\ref{sec:sims}, which corresponds to a satellite of $\sim 4
\times 10^8 \sm$ in stars.

We now focus on the metallicity distribution of the stars located in
the overdense segment of eccentricity identified previously. We have
divided the eccentricity interval [0.3, 0.5) into four small
intervals. The metallicity distribution obtained is shown in
Fig.~\ref{fig:feh_ecc}. The top panel corresponds to stars located
near the lower limit, while the high-eccentricity end is shown in
bottom right panel. Note that the metallicity distribution of the
stars in these two groups is very different. The stars with
eccentricities close to $\epsilon \sim 0.3$ are predominantly more
metal--rich, their distribution peaks around [Fe/H] $\sim - 0.4$
dex. On the contrary, stars with high orbital eccentricity $\epsilon
\sim 0.45$, are more metal--poor on average. Note as well, that there
seems to be a third peak in metallicity [Fe/H] $\sim -0.6$ for stars
with intermediate eccentricities, as depicted in the top right panel.
It is clear that in all eccentricity ranges, there is a contribution
from two or more groups of stars with similar metallicities. However,
the relative contribution of these groups varies as the eccentricity
changes. The fact that the characteristic metallicity of the stars in
this region of the APL-space varies with orbital eccentricity (in a
discontinuous fashion) leads us to conclude that more than one
disrupted galaxy has deposited debris in this region of phase-space.
Note that the single satellite case would imply a narrow distribution
of [Fe/H], which is independent of eccentricity, and is not consistent
with Fig.~\ref{fig:feh_ecc}.

Debris from disrupted galaxies is expected to have a somewhat extended
distribution of eccentricities, as shown in
Fig.~\ref{fig:arct_apl}. This fact together with the
above analysis, suggests that it is better to identify the various
satellites using the metallicity distribution of the stars in this
region of phase-space, rather than to do so on a purely dynamical
basis.

We explore further this line of thought by separating the set of stars
in this region of the APL space into three metallicity groups,
according to the characteristics of the histograms shown in
Fig.~\ref{fig:feh_ecc}. We focus on three groups with [Fe/H]~$\ge
- 0.45$ (Group 1; 120 stars), $-0.7 \le $~[Fe/H]~$ < -0.45$ (Group 2; 86
stars) and $-1.5 \le $~[Fe/H]~$< -0.7$ (Group 3; 68 stars), which
roughly correspond to the peaks/bumps seen in the right-hand side
panel of Fig.~\ref{fig:feh}.

\begin{figure}
\hspace*{-0.5cm}
\includegraphics[width=10cm]{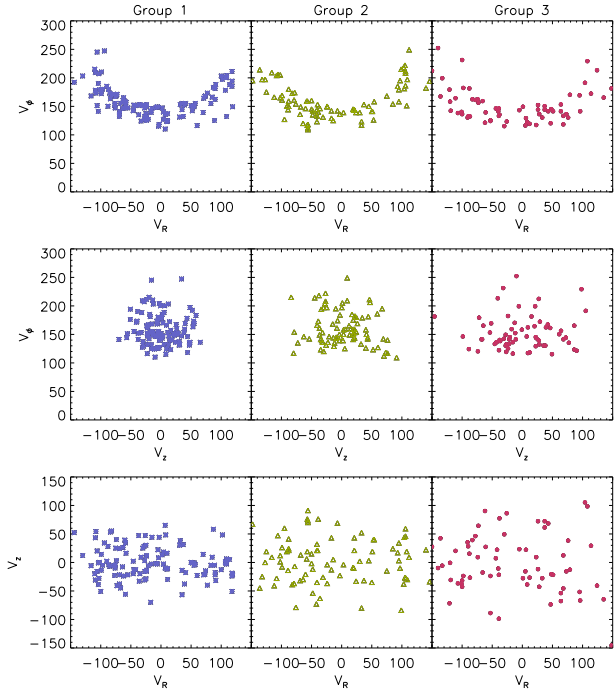}
\caption{Velocities of stars (in \kms) associated with the groups shown in
Fig.~\ref{fig:apl_groups}. Note that the velocity distribution in the
$z$-direction (second and third panels) is very different for
each group. Compare to the velocity distribution of satellite debris
shown in Figure \ref{fig:arct_vel}.}
\label{fig:vel_groups}
\end{figure}

Figure \ref{fig:apl_groups} shows the distribution of stars in this
overdense region of the APL-space, colour-coded according to their
metallicity. Note that there is considerable overlap between the
different groups, but that some are more dominant in certain regions
of phase-space (as discussed before). For example, the more metal-rich
stars preferentially have higher $L_z$ for a given apocentre. Figure
\ref{fig:vel_groups} shows their velocity distribution. The ``banana''
pattern in the upper panels is the same as that seen in
Fig.~\ref{fig:arct_vel} for satellite debris.  Note that the more
metal-rich stars (in blue; left panels) tend to have significantly
smaller $z$-velocities. In fact the $z$-velocity dispersion decreases
with metallicity. For Group 1: $\sigma_z \sim 28$ \kms, for Group 2:
$\sigma_z \sim 39 $ \kms, while for Group 3: $\sigma_z \sim 52$
\kms. While a similar trend of increasing vertical velocity dispersion
with metallicity exists for thin disk stars, it is by no means
comparable to the abrupt changes seen for this set of stars. As a
comparison, if we consider stars in the N04 sample with $\epsilon \le
0.2$ and group them according to the same three metallicity bins, the
$z$-velocity dispersions obtained are 16, 19 and 19.5 \kms
respectively, i.e. almost independent of [Fe/H]. For higher
eccentricity stars, those with $0.2 \le \epsilon \le 0.3$, $\sigma_z$
is 23, 30, 32 \kms in the same three metallicity bins, again
indicative of small variations with [Fe/H]. Therefore the strong
dependence of the $z$-velocity dispersion with metallicity found for
the stars in the overdense segment of constant eccentricity of the
APL-space is unique to this special set of stars. It constitutes
another manifestation of the particular nature of the stars found in
this region of phase-space.

\subsubsection{Age distribution}

Let us now focus on the age distribution of the stars in Groups 1, 2
and 3. The most reliable way to determine the age pattern of these
stars is by means of their HR diagram. We prefer not to use the
individual ages derived by N04 because of the large error bars. The
determination of individual ages for stars is subject to large
uncertainties, and it is more reliable to try to establish whether the
stars in a given group are mostly young or old, and to obtain a rough
estimate of their ages using isochrones. Since binary stars generally
introduce more scatter in the HR diagram, we will not include these in
the analysis. The identification of binaries has been done
spectroscopically and visually (see Sec.~3.2.4 of N04).

The uncertainty in the location of a star in the HR diagram comes from
two sources. Since the average error in the temperature is small (of
the order of 50 K to 100 K), it is mostly driven by uncertainties in
the absolute magnitude, which arise from the distance
determination. On average, the relative distance errors are in the
10\% - 13\% range. This implies an uncertainty in the absolute
magnitude of $\sim 0.2$~mag.

\begin{figure}
\hspace*{-1cm}
\includegraphics[width=9.5cm]{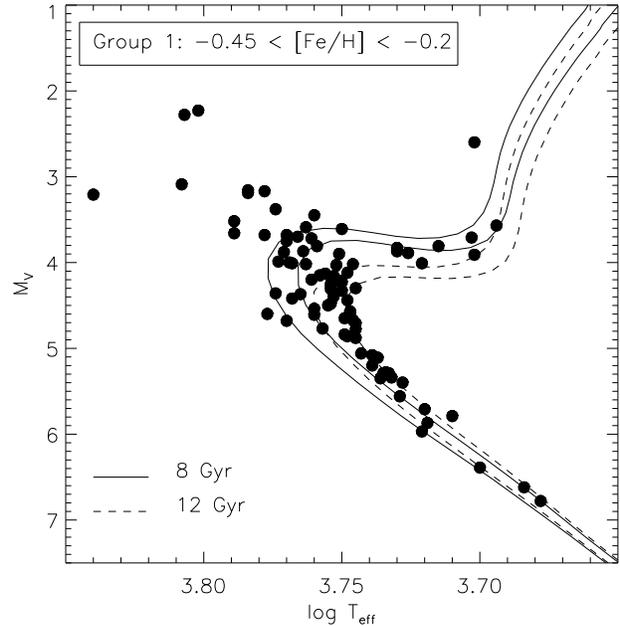}
\caption{HR diagram for the 88 single stars in Group 1. The
uncertainty in the location of a star in this diagram is $\sim 50 -
100$~K in the $x$-direction and $0.2$ magnitudes in the $y$-direction,
that is roughly twice the symbol size. Overplotted are two sets of
isochrones with metallicities close to the maximum (right-most) and
minimum values (left-most) found in this group, and for an 8 Gyr old
(solid) and a 12 (dashed) Gyr old population. In all cases, the
isochrones correspond to an $\alpha$-enhanced population, with
[$\alpha$/Fe] = 0.4 dex.}
\label{fig:cmd_s1}
\end{figure}

In Figure \ref{fig:cmd_s1} we plot the HR diagram of the 88 single
stars in Group 1.  It shows some very distinct features: $i$) there
are very few young stars; $ii$) there is evidence of two turn-off
points around $M_V \sim 3.7$ and $M_V \sim 4$; $iii$) there is a
clearly demarcated subgiant branch region. To establish whether and
how these features are linked, we have explored the Yonsei--Yale
single stellar population
library\footnote{http://www-astro.physics.ox.ac.uk/$\sim$yi/yyiso.html}
by Yi et al. (2001), and last updated by Demarque et al. (2004). In
Figure \ref{fig:cmd_s1} we overplot the sets of isochrones that best
match the various features as judged by eye-ball inspection. Two
isochrones correspond to a low age of $8$ Gyr (solid curves), that fit
well the location of the brightest TO point. Each of these two
isochrones is for a different metallicity that is close to the maximum
and minimum values found in Group 1: $-0.5$ dex and $-0.25$ dex,
respectively. A second set of (older) 12 Gyr isochrones (dashed
curves) reproduces well the location of the fainter TO point, and
yields an almost perfect match to the location of the stars on the
upper main sequence.

We can obtain an estimate of the relative importance of the two
populations by focusing on the region delimited by $3.5 \le M_V \le
5.5$, where the 8 Gyr and 12 Gyr isochrones show the least overlap. We
shall say that a star belongs to the 8 (12) Gyr population if it is
found to the left (right) of the most-metal rich 8 Gyr isochrone.  Out
of the 73 stars located in this region, 25 can be associated with the
younger isochrone, while the remaining 48 appear to be
older. Therefore, the ratio of the young to the old population in
Group 1 is roughly 1:3 versus 2:3 with respect to the total number of
stars in this group. It is worth noting that the kinematics of these
two populations are indistinguishable from one another.

\begin{figure}
\hspace*{-1cm}
\includegraphics[width=9.5cm]{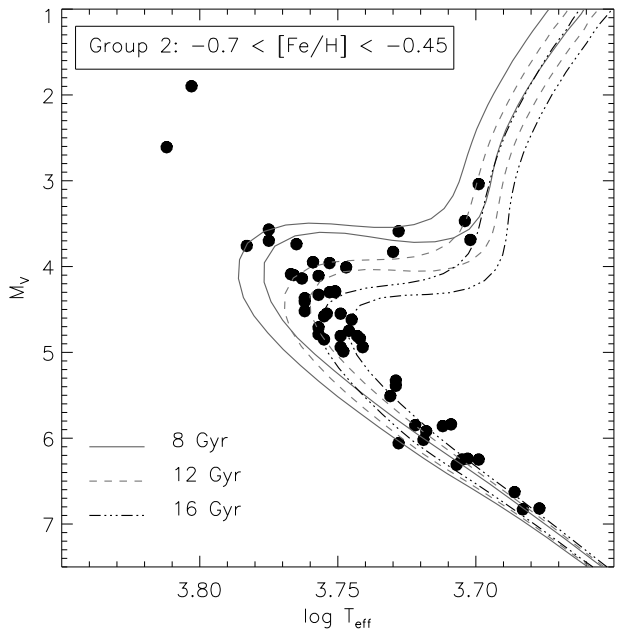}
\caption{HR diagram for the single stars in Group 2. Overplotted are
three sets of isochrones with metallicities close to the maximum
(right-most) and minimum values (left-most) found in the group, and
for [$\alpha$/Fe]~$= 0.4$ dex. The 8 Gyr (solid) reproduces well the
location of the TO point around $M_V \sim 3.6$. The intermediate 12
Gyr isochrone (dashed) matches the second TO point at $M_V \sim 4.5$,
while an older 16 Gyr isochrone appears to be required to fit the
location of the stars around $M_V \sim 4.7$ and $\log T_{eff} \sim
3.74$. Compare with the HR diagrams of Group 1 and 3 shown in
Fig.~\ref{fig:cmd_s1} and Fig.~\ref{fig:cmd_s3}, respectively.}
\label{fig:cmd_s2}
\end{figure}
\begin{figure}
\hspace*{-1cm}
\includegraphics[width=9.5cm]{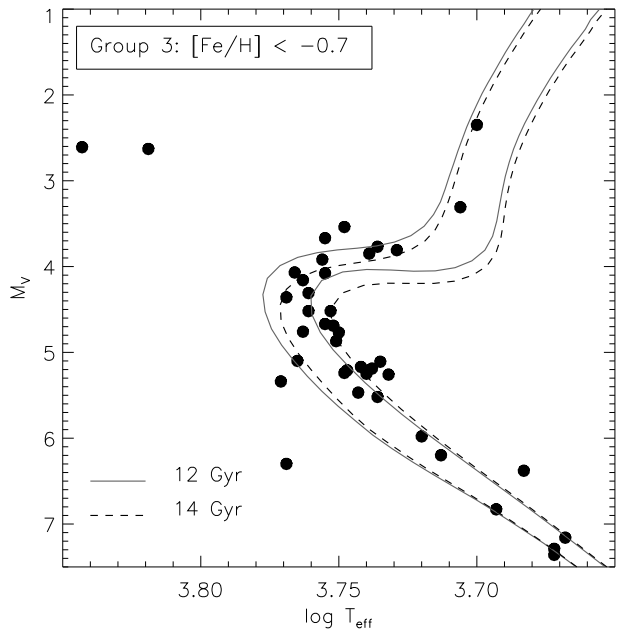}
\caption{HR diagram of all single 41 stars in Group 3. Overplotted are
two sets of isochrones with similar metallicity as the maximum
(right-most) and minimum values ([Fe/H]$\sim -1$ dex; left-most) and to
12 Gyr (solid) and 14 (dashed) Gyr of age, for [$\alpha$/Fe]~$=0.4$
dex. Unlike for groups 1 and 2, the stars in group 3 are consistent
with having all the same age, around 14 Gyr.}
\label{fig:cmd_s3}
\end{figure}

Figure \ref{fig:cmd_s2} shows the HR diagram for the 57 single stars
in Group 2. Like in the case of Group 1, few young stars are present,
and the subgiant and the bottom of the red giant branches can be
clearly identified. Furthermore, besides the two TO points visible
around $M_V \sim 3.6$ and $M_V \sim 4$, there is possibly a third
turn-off point at $M_V \sim 4.5$. We have overplotted the 8 Gyr old
and the 12 Gyr old isochrones for two metallicities [Fe/H]~$ = -0.75$
dex and [Fe/H]~$= -0.5$ dex, which are close to the maximum and
minimum values found in this group. Note that none of the 8 Gyr and 12
Gyr old isochrones provides a good match to the group of stars with
$M_V \sim 4.7$ and $\log T_{eff} \sim 3.74$. This is why we have also
plotted a third set of isochrones corresponding to an age of 16 Gyr which fit
the features better.  We estimate the relative importance of the
populations associated with these three sets of isochrones by focusing
on the region in the HR diagram defined by $3.5 \le M_V \le 5.5$. Out
of the 41 single stars located here, 6 can be associated with the 8
Gyr isochrones ($\sim 15$\%), 15 stars to the 12 Gyr old ($\sim 36$\%)
and the remaining 20 stars ($\sim 49$\%) to the 16 Gyr old isochrones.

In Figure \ref{fig:cmd_s3} we show the HR diagram of Group 3. In this
case, the dearth of young stars is even more striking than for Groups
1 and 2. Like before, the subgiant and red giant branches are quite
prominent. However, only one TO point can be clearly identified for
Group 3, around $M_V \sim 4.3$. This TO is fainter than what was found
for Group 1 and is slightly brighter than the weakest one in Group
2. For completeness, we have overplotted two sets of isochrones for 12
Gyr and for 14 Gyr, but the overall features of the HR diagram seem to
be well reproduced by the older 14-Gyr isochrone.

In our exploration of the stellar population libraries, we have found
that better matches to the location of the features (especially the
main sequence) in the various HR diagrams could be obtained for
$\alpha$-enhanced populations, in particular those with
[$\alpha$/Fe]~$=0.4$. This $\alpha$-enhancement is of similar
magnitude to that found in Galactic halo stars, although it occurs at
a relatively high-metallicity ([Fe/H]~$\ge -1$ dex) compared to the
halo. This result would suggest that the stars in Groups 1, 2 and 3
were formed from material that had been previously enriched both by
type I and type II supernovae, but predominantly by the
latter. Undoubtedly spectroscopic abundances are required to confirm
this speculation.

We now summarise the discoveries made in the last two Sections. Recall
that the dynamical analysis of the APL-space suggested the presence of
an excess of stars located in a segment of constant eccentricity, with
$0.3 \le \epsilon < 0.5$. In Sec.~\ref{sec:gral} we separated these
stars into Groups 1, 2 and 3 on the basis that the metallicity
distribution was varying with eccentricity in a discontinuous fashion
within this segment (Fig.~\ref{fig:feh_ecc}). We then found that these
three groups of stars also have slightly different kinematics,
particularly in the vertical ($z$) direction
(Fig.~\ref{fig:vel_groups}). Finally, we showed that the age
distribution of the stars varies from group to group, a point that is
clearly visible from simple inspection of the respective HR diagrams
(Figs.~\ref{fig:cmd_s1}, \ref{fig:cmd_s2} and \ref{fig:cmd_s3}). Using
single stellar population models, we quantified these differences, and
found evidence that Group 1 is formed by two populations of 8 and 12
Gyr of age (33\% and 66\% respectively); Group 2 by three populations
of 8 Gyr (only 15\%), 12 Gyr (36\%) and 16 Gyr (49\%), and Group 3 is
consistent with just one 14 Gyr old population. Therefore, the three
groups we identified are distinct in metallicity, kinematics as
well as in their age distribution.

\section{Summary and conclusions}
\label{sec:concl}

We have analysed the Geneva-Copenhagen survey of nearby, predominantly
disk, stars. This is the largest compilation of stars with accurate
and complete spatial and kinematic information, and for which
metallicities (derived from Str\"omgren photometry) are available
(99\% of the sample). We started this project with the goal
of quantifying the amount of substructure associated to mergers that the
Milky Way galaxy experienced in its evolution.

To guide the search for the remains of past minor mergers, we analysed
numerical simulations of the disruption of satellite galaxies. We
focused first on the time-independent regime (satellite + static host
galaxy), and secondly on the cosmological simulations of the formation
of disk galaxies by Abadi et al. (2003a). In both cases, we found that
substructure is present in the space defined by apocentre, pericentre
and $z$-angular momentum: the APL-space. While both apocentre and
pericentre depend on the functional form of the galaxy's gravitational
potential, perfect knowledge is not required for the substructure to
become apparent. Furthermore, such correlations are not significantly
altered by orbital evolution due to changes in the mass distribution
or to dynamical friction. It is true, however, that stars released in
different perigalactic passages have slightly different orbital
properties, and hence will generally be located in several smaller
lumps in the APL-space. However, these lumps are located along a
segment of constant eccentricity, thereby permitting the assessment of
a common origin.  Furthermore, we found that the substructure present
in the APL-space remains coherent for many Gyrs, well after the merger
has fully mixed.

The APL-space for the N04 catalogue shows large amounts of
substructure. The most prominent structures are related to the
superclusters Hyades-Pleiades, Sirius and Hercules, and are most
likely due to dynamical perturbations induced by the spiral arms and
the Galactic bar (Famaey et al. 2005). These structures are composed
by stars on disk-like orbits with relatively low eccentricity. The age
distributions of the HyPl and the Sirius superclusters are
indistinguishable from that of the full N04 sample, while Hercules
appears to have a more prominent old population. On the other hand,
the metallicity distributions of the stars in these superclusters are
very similar and typical for the thin disk, peaking at [Fe/H] $\sim
-0.13$ dex for Hercules, $-0.08$ dex for HyPl and $-0.18$ dex for
Sirius with a dispersion of $\sim 0.2$ dex in all cases.


The detailed statistical analysis of the APL-space reveals the
presence of the order of ten other overdensities at significance
levels higher than 99\%. Unlike the case of the superclusters, these
overdensities are located along two to three segments of constant
eccentricity, as predicted for substructures that are the result of
minor mergers. There are 274 stars in this region of the APL-space,
which is delimited by eccentricity $0.3 \le \epsilon < 0.5$. The
metallicity distribution of these stars is inconsistent with that of
the whole N04 sample at the $10^{-8}$ level, as quantified by a
KS-test. Furthermore, the HR diagram shows unambiguously that these
stars are predominantly old; prominent subgiant and red giant branches
are easily distinguishable, and the very few young stars present are
put in evidence by a turn-off point around $M_V \sim 3.8 - 4$.  All
this strongly supports the hypothesis that these substructures
represent the remains of satellite galaxies that merged with the Milky
Way several Gyrs ago.

The metallicity distribution of the stars in this overdense region of
the APL-space varies with eccentricity in a discontinuous
fashion. This allows the separation of these stars into three Groups,
which we believe represent at least three accreted galaxies. These
three groups of stars are dissimilar not only in their metallicity
distribution, but also have different kinematics in the vertical ($z$)
direction and distinct age distributions. The most metal-rich group
([Fe/H]~$>-0.45$ dex) has 120 stars which are distributed into two
stellar populations of 8 Gyr (33\%) and 12 Gyr (67\%) of age, as
deduced by comparison to the isochrones by Yi et al. (2001) and from
the presence of two turn-off points at $M_V \sim 3.7$ and $M_V \sim
4$. The second Group with $\langle$[Fe/H]$\rangle$~$\sim -0.6$ dex has
86 stars, and is constituted by three populations of 8 Gyr (15\%), 12
Gyr (36\%) and 16 Gyr (49\%). Finally, the third Group has 68 stars,
with typical metallicity around $-0.8$ dex, and which appear to be
part of a single 14 Gyr old population.  Whereas none of the Groups
was identified as such previously, there seems to be considerable
overlap between Group 2 and the Arcturus ``stream'' (Navarro et al
2004), both in their kinematics as in their metallicities.

The characterization of the progenitor galaxies, either in terms of
their initial mass or their stellar properties is possible in this
case thanks to the multi-faceted nature of the N04 sample.  A direct
comparison to the numerical simulation presented in
Sec.~\ref{sec:sims} suggests that the accreted satellites had an
initial mass $\sim 4 \times 10^8 \sm$ (much lower masses can be
excluded on the basis of their extent in the APL-space). The
relatively high-metallicity characteristic of these groups is
consistent with the expected correlation between luminosity (or mass)
and [Fe/H] found in present-day galaxies (Mateo 1998). Further insight
into the properties of these building blocks could be gained from
high-resolution spectroscopy. Knowledge of the detailed abundance
patterns could prove to be extremely useful as a way of confirming the
origin and the history of the different groups (Freeman \&
Bland-Hawthorn 2002).

The substructures identified cannot be uniquely associated to a single
traditional Galactic component, since they are in a transition region
that overlaps with both the thin and the thick disks. This assessment
is based on the metallicity distribution of their stars (peak values
in the range $-0.4$ dex to $-0.8$ dex) as well as on their vertical
velocities ($\sigma_z \sim 28 - 52$ \kms). The identification of
debris in the Galactic disks whose origin can be traced back to more
than one satellite galaxy provides support to the accretion scenario
for the formation of the thick disk (Abadi et al.~2003b). However, it
is not clear how this scenario differs from the dynamical heating of a
previously existing thin disk by (in this case, more than) one minor
merger. In this context, a natural next step consists in establishing
how the properties of a pre-existent thin disk differ dynamically from
those of accreted satellites (Villalobos \& Helmi, in preparation).

Our findings suggest that the peak in the merging activity that led to
the Milky Way has taken place before $z \sim 1$ for a $\Lambda$CDM
cosmology, with little accretion of (relatively) massive galaxies
after that. This is not unlike the results of the simulations of Abadi
et al. (2003b).

An important issue raised by this study concerns the traditional
separation of the Galaxy into distinct components, and its description
in terms of a set of coarse-grained distribution functions that
satisfy Jeans theorems (Binney \& Tremaine 1987). The Galactic disk
appears not to be in steady-state, i.e. its phase-space density is
probably changing in time as a result of the perturbations produced by
spiral density waves and the Galactic bar. Moreover, we found that the
fine-grained distribution function still contains substantial
discernible information about the evolutionary history of the
Galaxy. It will be essential to establish how to optimally
characterize the phase-space structure of the Galaxy taking into
account its graininess, particularly in the context of billion-star
surveys such as Gaia (Binney 2005).

Galactic structure has come to be a very complex field. The vast
datasets that will become available in the near future, such as RAVE
(Steinmetz 2003), SDSS-II with SEGUE and ultimately Gaia (Perryman et
al. 2001) will drive this to a summit.  However, access to such
complexity and richness is the key to disentangle the dynamical and
evolutionary history of the Milky Way. Through studies like that
presented here, we should hope to eventually be able to reconstruct
the star formation, chemical enrichment pattern and evolutionary
history of the building blocks of galaxies.

\section*{Acknowledgments}
Eline Tolstoy is gratefully acknowledged for very stimulating
discussions. The anonymous referee is thanked for constructive
comments.  We have made use of the Besancon model of the Galaxy, and
are very grateful to Annie Robin and her collaborators for making such
a useful tool accessible to the community. This work has been
partially financed by the Netherlands Organization for Scientific
Research (NWO). JFN acknowledges support from the Alexander von
Humboldt Foundation and from the Leverhulme Foundation. BN gratefully
acknowledges substantial financial support from the Carlsberg
Foundation, the Danish Natural Science Research Council, the Swedish
Research Council, the Nordic Academy for Advanced Study and the Royal
Physiographic Society in Lund.

\label{lastpage}
\end{document}